\documentclass[12pt]{article}

\usepackage[margin=1in]{geometry}
\usepackage{graphicx}
\usepackage{amsmath}
\usepackage{amssymb}
\usepackage{booktabs}
\usepackage{url}
\usepackage{cite}
\usepackage[T1]{fontenc}
\usepackage[utf8]{inputenc}
\usepackage[english]{babel}
\usepackage{siunitx}
\usepackage{float}
\usepackage{arydshln}

\newcommand{\hlsfml}{\texttt{hls4ml}}
\newcommand{\snntorch}{\texttt{snnTorch}}
\newcommand{\pytorch}{\texttt{PyTorch}}
\newcommand{\apfixed}[2]{\texttt{ap\_fixed<#1,#2>}}
\newcommand{\apfixeds}[2]{\texttt{<#1,#2>}}

\begin{document}

\title{Spiking neural networks for streaming qubit readout}

\author{Barry M. Dillon$^1$, Aqib Javed$^1$, Jim Harkin$^1$,\\Patryk D\k{a}bkowski$^{2,3,4,5}$, Benjamin Lienhard$^{2,3,4}$\\
{\small $^1$ ISRC, Ulster University, Derry, BT48 7JL, Northern Ireland}\\
{\small $^2$ Technical University of Munich, Department of Physics, 85748 Garching, Germany}\\
{\small $^3$ Walther-Meißner-Institut, Bayerische Akademie der Wissenschaften, 85748 Garching, Germany}\\
{\small $^4$ Munich Center for Quantum Science and Technology, 80799 Munich, Germany}\\
{\small $^5$ Zurich Instruments, 8005 Zürich, Switzerland}\\
\texttt{\small b.dillon@ulster.ac.uk}}

\date{}
\maketitle

\begin{abstract}
Fast and accurate qubit-state assignment is essential for feedback, calibration, and error correction in quantum processors.
In superconducting platforms, frequency-multiplexed readout makes this task intrinsically multivariate as measured traces can encode crosstalk, qubit-state relaxation events, and other transient nonidealities that are not fully captured by conventional matched filtering. 
Here, we introduce spiking neural network discriminators for superconducting qubit readout. 
By processing the measurement window in successive time chunks, the networks exploit temporal structure and update classification scores as data arrive, rather than waiting until the end of the readout window.
The spiking networks outperform matched-filter discrimination and approach the accuracy of a full-trace artificial neural network. 
Beyond reaching the performance of artificial neural networks, the key advantage of SNNs is that they provide a streaming, time-resolved estimate of the qubit state that evolves as the readout signal is acquired.
Using quantisation-aware training and \hlsfml{} synthesis, we further demonstrate that each FPGA inference update can be completed before the next readout chunk arrives. These results establish spiking neural networks as a promising route to low-latency, real-time qubit readout on FPGA hardware, with broader implications for time-critical quantum-control and scientific-inference applications.
\end{abstract}

\noindent\textit{Keywords}: superconducting-qubit readout; spiking neural networks; FPGA inference;\\
quantisation-aware training; hls4ml

\section{Introduction}

Scaling quantum computation from a few tens to tens of thousands of qubits promises new capabilities in science and technology, from more accurate simulations of molecules and materials to advances in chemistry, drug discovery, and complex optimisation~\cite{montanaro2016quantum}. Realising this promise requires suppressing and managing errors through quantum error correction (QEC) and error mitigation, both of which rely on repeated, high-fidelity access to the processor~\cite{PhysRevA.52.R2493,Kitaev_1997,Campbell2017}. Readout, the measurement step that converts quantum information into a classical record, is therefore not merely a final diagnostic operation, but a central component of scalable quantum computation.

For many quantum-computing platforms, readout remains among the most error-prone and time-consuming operations. This is particularly consequential in feedback-based protocols, calibration routines, and QEC cycles, where decisions must be made rapidly enough to act before encoded information changes~\cite{cramer2016repeated}. Improving readout, therefore, requires addressing two coupled objectives: increasing state-assignment accuracy while reducing the latency between measurement acquisition and classification.

Superconducting qubits are among the leading platforms for realising useful quantum processors. They combine fast gate operations with a mature ecosystem for microwave control and measurement, which naturally supports parallel, frequency-multiplexed readout~\cite{10.1063/1.5089550}. These features make superconducting architectures well-suited to QEC codes and scalable quantum hardware~\cite{Acharya2025}. At the same time, superconducting devices remain vulnerable to hardware imperfections and environmental noise. In multiplexed readout, the measured signals can contain crosstalk, relaxation events, and transient nonidealities, making qubit-state assignment an intrinsically time-dependent and multivariate inference problem~\cite{Lienhard:2022tiq}.

A common approach to qubit-state assignment in superconducting circuits is matched filtering (MF)~\cite{PhysRevA.91.022118}. In this method, readout traces are acquired for qubits prepared in known initial states and used to construct state-dependent templates. A new measurement record is then classified according to its overlap with these templates. This procedure typically requires acquiring the full trace before computing the MF response.

MF is simple and effective, but it has important limitations. Because it compresses the measurement record into a fixed overlap with a predetermined template, it is not generally optimised to detect rare or time-localised events that may occur at different points during the readout window, such as relaxation, transient crosstalk, or other nonideal dynamics~\cite{Lienhard:2022tiq}. Moreover, depending on the preprocessing required for the readout traces, MF evaluation can become computationally costly in latency-constrained settings.

To improve both readout accuracy and latency, machine-learning methods have been developed for superconducting-qubit-state assignment~\cite{Lienhard:2022tiq}. Artificial neural networks (ANNs), for example, can be trained directly on measurement records from qubits prepared in known states, learning a nonlinear map from the readout trace to the corresponding state assignment. More broadly, the literature on machine-learning-assisted readout includes Bayesian approaches~\cite{Cosco:2023jax}, reservoir computing~\cite{Kent:2025fxy}, and hybrid methods that combine neural networks with high-level MF-based observables~\cite{Maurya:2022ida}. We discuss this body of work in more detail in Sec.~\ref{sec:readout-traces}.

ANNs can improve both readout accuracy and inference latency when deployed on FPGA hardware~\cite{Reuer:2022efh}. In conventional implementations, however, the network is evaluated only after the full readout pulse has been acquired and preprocessed. The total readout latency is therefore the sum of the pulse duration and the network-inference latency. As FPGA-implemented neural networks can now achieve inference times on the order of \SI{50}{\nano\second}, the dominant latency bottleneck is no longer the network evaluation itself, but the need to wait for the complete readout record.

\begin{figure}[!htbp]
\centering
\includegraphics[width=0.85\linewidth]{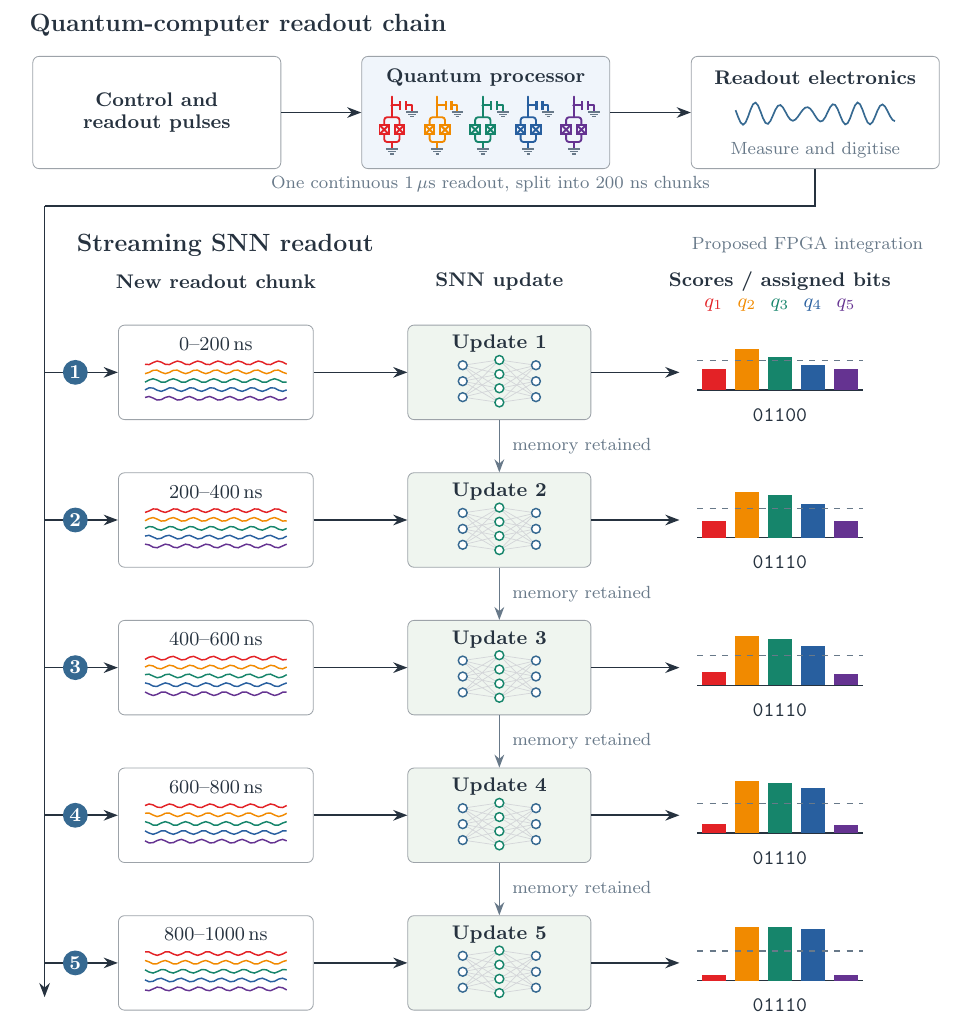}
\caption{This figure depicts the real-time updates of the SNN measuring qubits in the $01110$ state over $200$ns timesteps. The coloured lines represent the demodulated traces for each qubit.
These are then averaged and fed directly to the SNN, which updates its internal neuron states and predictions for the qubit readout classes.}
\label{fig:diagram-snn}
\end{figure}

In this work, we propose using Spiking Neural Networks (SNNs) synthesised for FPGA hardware to achieve high-accuracy, low-latency readout. 
The SNN discriminators process the readout signal sequentially as it arrives, rather than waiting until the readout has finished.
This allows the classification process to start earlier, making better use of the available resources in a severely resource-limited setting.
It also opens the possibility that time-dependent features could contribute to the classification before the full measurement window has elapsed. 
The idea is depicted schematically in Fig.~\ref{fig:diagram-snn}.
This approach provides a natural route to low-latency inference while retaining sensitivity to transient structures that conventional readout methods lack. 
Our results show that spiking neural networks can improve superconducting-qubit state assignment and provide a hardware-compatible path toward real-time readout for feedback, calibration, and quantum-error-correction workflows.

SNNs are well-suited to this task because they are both inherently temporal and are well-suited to quantisation and deployment in resource-limited settings.
SNNs replace the continuous, stateless activations of ANNs with event-like spikes and an internal dynamical state that evolves over time~\cite{ANDERSON1972197,Maass:1997spiking}.
Here, we exploit this temporal memory to process successive chunks of the readout trace and update the qubit-state assignment throughout the acquisition.
Recent advances in SNN synthesis for FPGA hardware using \hlsfml{}~\cite{dillon2026spikingneuralnetworkinference} also enable realistic estimates of latency and resource requirements for deployment.

Together, these results establish streaming SNN-based discrimination as a neuromorphic route to low-latency qubit-state readout. 
By updating state assignments during acquisition, the method reduces the dependence of decision latency on the full readout-pulse duration while retaining sensitivity to transient features such as relaxation and crosstalk. 
Its combination of accuracy, low latency, and FPGA-compatible implementation addresses a key requirement for feedback, calibration, and QEC in superconducting quantum processors. More broadly, this work highlights neuromorphic architectures as real-time inference engines for time-critical quantum technologies.

\section{Superconducting Qubits}
\label{sec:superconducting-qubits}

Superconducting qubits encode quantum information in the lowest levels of weakly anharmonic Josephson-junction circuits and are typically measured dispersively through qubit-state-dependent shifts of microwave resonators. In the multiplexed-readout dataset introduced by Lienhard et al.~\cite{Lienhard:2022tiq}, the experimental platform was a five-qubit transmon processor in which each qubit with transition frequencies between \SIrange{4.3}{5.2}{\giga\hertz} and lifetimes ranging from \SIrange{7}{40}{\micro\second} was coupled to an individual readout resonator in the range of \SIrange{7}{7.25}{\giga\hertz}. All five resonators were coupled to a common feedline for frequency-multiplexed transmission measurement. This architecture provides a compact but realistic benchmark for multiqubit readout: it combines simultaneous dispersive measurement, shared analogue signal paths, resonator-frequency crowding, amplifier noise, non-Gaussian IQ distributions, inter-channel correlations, and readout crosstalk.

The dataset consists of labelled single-shot measurements acquired after preparing the device in the full set of $2^5=32$ computational-basis states. For each shot, the prepared basis state serves as the class label, and a frequency-multiplexed microwave readout pulse containing tones near the five readout-resonator frequencies is applied. The reflected signal is amplified, downconverted, digitised, and stored as time-resolved in-phase and quadrature (IQ) traces. Thus, each sample contains the temporal IQ response of the frequency-multiplexed readout chain, rather than only an integrated scalar per qubit. This makes the dataset reusable for benchmarking discriminators that exploit either conventional integrated features, full-time-domain trajectories, correlations among readout channels, or hardware-efficient compressed representations.

The five transmons, individual dispersive readout resonators, shared frequency-multiplexed feedline, labelled preparation of all $32$ computational-basis states, and time-resolved IQ acquisition provide the experimental setting used throughout this work. The following section describes how these traces are processed and introduces the benchmark methods and performance metrics.

\subsection{Qubit-State Readout Traces}
\label{sec:readout-traces}

We restrict the analysis of the five-qubit frequency-multiplexed readout traces to the first \(1\,\mu\mathrm{s}\) of each acquisition.
The sampling period is \(\Delta t\!=\!2\,\mathrm{ns}\), yielding \(N\!=\!500\) samples per trace.
The in-phase (\(I\)) and quadrature (\(Q\)) samples oscillate at their respective intermediate frequency (IF) and are combined as
\begin{equation}
z[n] = I_{\mathrm{IF}}[n] + i Q_{\mathrm{IF}}[n],
\qquad n=0,\ldots,N\!-\!1 .
\label{eq:raw-iq}
\end{equation}
For qubit \(q\), the trace is demodulated at IF, \(f_q\),
\begin{equation}
x_q[n] = z[n]\exp\!\left(i2\pi f_q n\Delta t\right)
       = I_q[n] + iQ_q[n].
\label{eq:demod}
\end{equation}
After demodulation, each qubit trace is smoothed with a centred moving-average boxcar,
\begin{equation}
\tilde{x}_{q,W}[n] =
\frac{1}{W}\sum_{r=-\lfloor(W-1)/2\rfloor}^{\lceil(W-1)/2\rceil}
x_q[\mathrm{clip}(n+r,0,N\!-\!1)].
\label{eq:boxcar}
\end{equation}
For the figures showing example traces, we use $W\!=\!5$, while for the matched-filter benchmark, we use $W\!=\!25$.
Fig.~\ref{fig:readout-example} shows a representative digitally-unmodulated trace and the corresponding demodulated trace, where we have used the $W\!=\!5$ boxcar on the latter.
Fig.~\ref{fig:readout-average} represents the average demodulated response for the prepared ground and excited states, which also uses the $W\!=\!5$ boxcar.
The graph reveals that, in all cases, there are clear separations between the ground and excited states in at least one channel, except for $q_2$, where the separation is more subtle.

The performance metric used in this work is state-assignment fidelity, i.e. the probability that the predicted assignment is correct.
For qubit \(q\), the single-qubit assignment fidelity is
\begin{equation}
F_q = \frac{1}{2}\left[
P(\hat b_q\!=\!0 \mid b_q\!=\!0) + P(\hat b_q\!=\!1 \mid b_q\!=\!1)
\right],
\label{eq:fq}
\end{equation}
where \(b_q\) is the prepared bit and \(\hat b_q\) is the assigned bit.
It represents the quality of the qubit readout, where a value of $1$ would imply a perfect classification of the qubit state.
Since we consider multiqubit readout traces, we focus most of our attention on the geometric mean of the five scores,
\begin{equation}
F_{\mathrm{geom}} = \left(\prod_{q=1}^{5}F_q\right)^{1/5}.
\label{eq:fgeom}
\end{equation}
Another possibility is to use the accuracy, defined as the fraction of samples for which all five bits are assigned correctly.
However, this is much more sensitive to outliers, since an assignment that is incorrect by one bit flip is considered just as incorrect as one that is incorrect in all five bits.
To diagnose readout crosstalk, we use the cross-fidelity matrix
\begin{equation}
F^{\mathrm{CF}}_{ij}
= 1
- P(\hat b_i\!=\!1\mid b_j\!=\!0)
- P(\hat b_i\!=\!0\mid b_j\!=\!1),
\label{eq:cross-fidelity}
\end{equation}
where \(i\) indexes the assigned qubit and \(j\) indexes the prepared qubit.
The diagonal gives the single-qubit discrimination contrast, \(F^{\mathrm{CF}}_{ii}=2F_i-1\) for balanced test labels, while off-diagonal entries quantify how the prepared state of one qubit biases the assignment of another.

\subsection{Matched Filters}

The matched filter (MF) is a good baseline method for this qubit discrimination task.
It uses the demodulated and boxcar-smoothed \(I,Q\) trace for each qubit.
For each qubit, the ground- and excited-state template means \(\mu_{q,0}\) and \(\mu_{q,1}\), and variances \(\sigma^2_{q,0}\) and \(\sigma^2_{q,1}\), are estimated from the training split.
For the MF estimation, we only use traces with the other four qubits prepared in the ground state\footnote{Although in the end, this had a very small effect on the results compared to allowing the other qubits to be in either state.}.
The filters are vectors with dimensions equal to the number of timesteps in the trace, and there is a filter for each (I,Q) trace of each qubit.
The Fisher-style kernel defines the filters as
\begin{equation}
h_q =
\frac{\mu_{q,1}-\mu_{q,0}}
{\sigma^2_{q,1}+\sigma^2_{q,0}+0.05\,m_q},
\qquad
h_q \leftarrow \frac{h_q}{\lVert h_q\rVert_2},
\label{eq:mf-kernel}
\end{equation}
where \(m_q\) is the median of \(\sigma^2_{q,1}+\sigma^2_{q,0}\).
For a test trace \(j\), the MF score is the dot product between the filter and the corresponding demodulated trace
\begin{equation}
s_{j,q} = \left\langle h_q, \tilde{x}_{j,q}\right\rangle .
\label{eq:mf-score}
\end{equation}
One threshold per qubit is chosen on the training split,
\begin{equation}
\theta_q^\star =
\underset{\theta}{\arg\max}\,
\frac{1}{2}\left[
P(s_{j,q}\le \theta\mid b_q=0) +
P(s_{j,q}> \theta\mid b_q=1)
\right],
\qquad
\hat b_{j,q}=\mathbb{I}[s_{j,q}>\theta_q^\star].
\label{eq:mf-threshold}
\end{equation}
The MF benchmark reaches \(F_{\mathrm{geom}}=0.8960\), with most of the loss coming from the second qubit.

\subsection{Readout Errors}

The main sources of readout error are noise in the measurement chain, qubit-state transitions during the acquisition, and crosstalk between simultaneously measured channels.
For a qubit prepared in the excited state, energy relaxation can occur before or during the readout pulse.
The resulting trace initially follows the excited-state response before moving towards the ground-state response.
An MF that integrates the complete trace into a single score can therefore place these events between the two main state distributions or incorrectly assign them to the ground state.
Time-resolved classifiers can instead identify the transient trajectory associated with relaxation~\cite{Martinez:2020bbx,Lienhard:2022tiq}.

Frequency-multiplexed readout introduces additional errors because several readout tones, resonators, and qubits share the same feedline and amplification chain.
Crosstalk can arise from insufficient spectral separation between resonators, parasitic coupling between qubits and neighbouring readout resonators, residual photon populations, and interactions between signals in the amplifiers, mixers, and digitisation electronics~\cite{Lienhard:2022tiq}.
These effects cause the measured response of one qubit to depend on the states of the other qubits, so the joint readout distribution is not generally separable into five independent classification problems.

Other nonidealities include imperfect state preparation, thermal excitation, amplifier noise, and measurement-induced state transitions.
Together, these effects produce broadened, distorted, and sometimes non-Gaussian distributions in the integrated (I,Q) plane.
A multiqubit classifier can learn correlations between channels and compensate for some of these system-dependent effects. At the same time, access to the time evolution of the traces provides information that is lost by complete-window integration.

\subsection{Machine-Learning Approaches}

Conventional superconducting-qubit readout pipelines typically integrate the IQ trace using boxcar or matched-filter weights and then classify each qubit independently using thresholds or Gaussian-style discriminators. These approaches are well matched to isolated single-qubit readout, but they become suboptimal for simultaneous multiqubit measurement because the joint measurement distribution is not generally separable across channels. In particular, multiplexed readout can produce correlated assignment errors and state-dependent distortions that are not captured by independent single-qubit decision boundaries~\cite{Lienhard:2022tiq}.

ML methods provide a natural extension of matched filtering because they can learn nonlinear decision boundaries directly from calibrated readout data.
Early work showed that support-vector machines and linear discriminant analysis can improve single-shot state assignment beyond simple thresholding~\cite{Magesan:2014bto}.
Subsequent approaches employed time-resolved models, such as recurrent neural networks~\cite{Flurin:2018rzr}, hidden Markov models~\cite{Martinez:2020bbx}, and reservoir computing methods~\cite{Angelatos:2020fvv,Kent:2025fxy}, to exploit the temporal structure of the measurement record.
For multiplexed superconducting readout, neural-network classifiers have also been shown to improve multiqubit assignment by learning correlations between channels and reducing state-dependent crosstalk~\cite{Lienhard:2022tiq,Kim:2021ixr}.
Other recent work has combined learned models with physically motivated readout features, Bayesian inference, calibration routines, and FPGA deployment workflows~\cite{Maurya:2022ida,Cosco:2023jax,Barrett:2023nhg,DiGuglielmo:2025zod}.

Lienhard et al. used fully connected feed-forward neural networks as multiqubit state discriminators operating on the multiplexed readout record. Trained and evaluated on the same five-qubit dataset used here, these networks improved multiqubit state assignment relative to contemporary conventional discriminators, achieving up to a 25\% reduction in assignment error and up to an order-of-magnitude reduction in readout crosstalk~\cite{Lienhard:2022tiq}. 

The main advantage of these methods is that they use more information than a single integrated score.
In particular, access to the time evolution of the trace can retain information about transient effects such as early relaxation. At the same time, multiqubit inputs allow the classifier to correct correlated readout errors.
Our SNN approach retains these advantages while processing the readout stream sequentially, allowing the assignment to be updated in real time and enabling early classification of qubit states.

\section{Spiking Neural Networks}
\label{sec:snns}

SNNs replace the continuous stateless activations of ANNs with spike-events and have an internal dynamical state that persists and changes in time~\cite{ANDERSON1972197,Maass:1997spiking}, and so are often described as the third generation of neural networks. 
They deviate from the traditional view of the ANNs and provide a much closer model of how the brain processes information.
Deep learning with SNNs has also proven successful~\cite{TAVANAEI201947,oconnor2016deepspikingnetworks}. It has generated a lot of interest in processing temporal data on hardware that achieves low power and latency for sparse, event-like spiking data. 
The development of dedicated neuromorphic processors such as TrueNorth~\cite{Merolla:2014million}, Loihi~\cite{Davies:2018loihi,shrestha2023efficientvideoaudioprocessing}, and SpiNNaker~\cite{Furber:2014spinnaker} has shown that the scalability of large-scale event-based computing systems is a realistic goal.  
In scientific domains, many instruments and experimental setups already use synchronous FPGAs for low-latency, real-time machine learning.  
SNN designs for deployment on FPGAs are digital, clock-driven systems that preserve stateful neuron dynamics and temporal readout and, in some cases, achieve lower power consumption due to sparse spiking activity.

In recent years, interest in SNNs for low-latency scientific applications has grown~\cite{mehonic2024roadmapneuromorphiccomputingemerging}.
Applications include temporal scientific data classification~\cite{10.1145/3183584.3183612}, radiation detection~\cite{10.3389/fphy.2024.1334298}, and vertex reconstruction in neutrino-nucleus interactions~\cite{8683736}.
At the Large Hadron Collider (LHC), SNNs have been studied for applications such as data filtering~\cite{kulkarni2023onsensordatafilteringusing}, particle tracking~\cite{Coradin_2025}, readout for hadron calorimeters~\cite{lupi2025neuromorphicreadouthadroncalorimeters}, and real-time anomaly detection~\cite{Dillon:2025yqe}.
In this work, we use the temporal state within the SNN to enable the real-time processing and updating of the qubit readout decision.

SNNs are constructed from spiking neurons.
The most widely used spiking neuron is the Leaky Integrate and Fire (LIF) neuron, described by
\begin{align}
    y_t &= \begin{cases}
0 & \text{if } u_t < u_{\text{thresh}} \\
1 & \text{if } u_t \geq u_{\text{thresh}}
\end{cases} \\
    u_{t+1} &= \beta u_t + w x_{t+1} - \beta y_t u_{\text{thresh}} + b.
\end{align}
This describes a neuron receiving an input current $x_t$ (at timestep $t$) contributing to its membrane potential $u_t$, i.e. the internal state of the spiking neuron.
Over time steps, spike events arrive and accumulate at the neuron.
The potential $u_t$ accumulates these current contributions while also leaking potential over time, with $0\leq\beta\leq1$ controlling the leakage rate.
Once the membrane potential $u_t$ exceeds the neuron's threshold potential $u_{\text{thresh}}$, the neuron generates a spike and the membrane potential resets.
The spiking action is not differentiable, but this nonlinearity gives SNNs their expressivity.

The spiking neuron is used in the same way as a ReLU neuron in an ANN, with layers of neurons separated by linear transformations parameterised by weights and biases.
To optimise the weights, biases, and parameters of the spiking neurons, we use surrogate-gradient methods.
A differentiable proxy for the spike derivative is used during backpropagation, while on the forward pass, the non-differentiable spiking dynamics are kept~\cite{oconnor2016deepspikingnetworks,Zenke_2018,8891809}.  
We use the \snntorch{} package for training the SNNs in this paper~\cite{Eshraghian:2023lessons}.
The \snntorch{} package exposes spiking neurons as \pytorch{}~\cite{paszke2019pytorchimperativestylehighperformance} modules and provides tools for training SNNs with modern deep-learning workflows.  

\subsection{Hardware Acceleration with \hlsfml{}}

A key part of this work is demonstrating that SNNs can perform the readout within the strict resource and latency requirements dictated by both the $\mu$s timescales on which quantum controllers operate and the timesteps we choose for our SNN readout.
For this, we use the \hlsfml{} package~\cite{Duarte:2018ite,Fahim:2021hls4mlCodesign,Schulte:2025hls4mlPlatform}, which is designed to map trained neural networks to HLS firmware for low-latency FPGA and ASIC applications.  
Recent work has extended the functionality to support SNNs~\cite{dillon2026spikingneuralnetworkinference}, which we use to translate trained SNN models into HLS code and obtain C-synthesis estimates with Vitis HLS.
For specific details on how SNNs are synthesised, including details on window-size logic, we refer the reader to~\cite{dillon2026spikingneuralnetworkinference}.

\section{SNN-based Qubit Readout}
\label{sec:snn-readout}

\begin{figure}[t]
\centering
\includegraphics[width=0.72\linewidth]{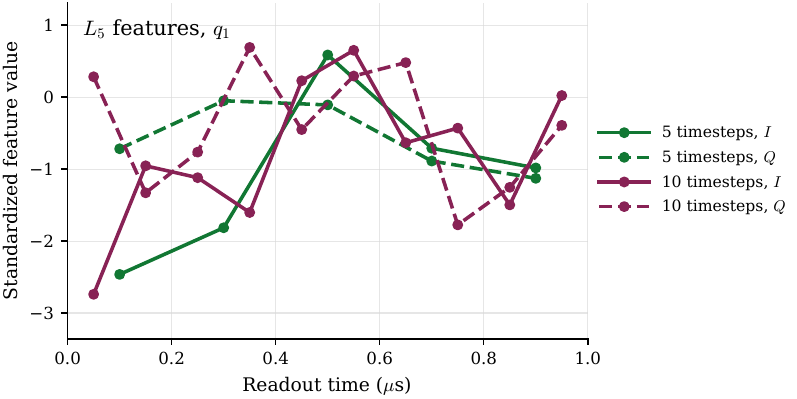}\\[0.6em]
\includegraphics[width=0.72\linewidth]{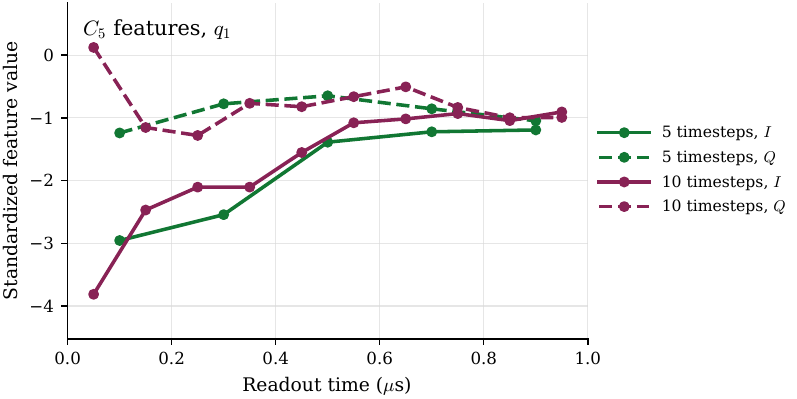}
\caption{Local \(L_5\) feature values (top) and cumulative \(C_5\) feature values (bottom) for \(q_5\), comparing $T\!=\!5$ and $T\!=\!10$ readout summaries over the \(1\,\mu\mathrm{s}\) acquisition window. Solid lines show the \(I\) component and dashed lines show the \(Q\) component.}
\label{fig:l5-c5-features}
\end{figure}

The sampling period for the readout traces is $2$ns, which is much too short to have the SNN read each update individually.
As the data streams in, an analogue-to-digital converter (ADC) will split the stream into $T$ chunks and compute the average for each chunk before passing the data chunk by chunk to the SNN.
We consider two options $T\in\{5,10\}$ for the number of chunks, corresponding to chunk durations of $200$ns and $100$ns, respectively.
For timestep \(t\), define \(M=N/T\) and write the chunked data stream as
\begin{equation}
\mathcal{S}_t = \{(t-1)M,\ldots,tM-1\}.
\label{eq:timestep-set}
\end{equation}
We then define the local feature vector as just the average of the $(I,Q)$ chunk for each qubit, forming a $10$-dim vector
\begin{equation}
L_5(t) =
\left[
\frac{1}{M}\sum_{n\in\mathcal{S}_t} I_{q,5}[n],~~
\frac{1}{M}\sum_{n\in\mathcal{S}_t} Q_{q,5}[n]
\right]_{q=1}^{5},
\label{eq:l5}
\end{equation}
where $I_{q,5}$ and $Q_{q,5}$ are the real and imaginary parts of $\tilde{x}_{q,5}$.
We also define a cumulative feature vector which sums all previous contributions to the traces,
\begin{equation}
C_5(t) =
\left[
\frac{1}{tM}\sum_{n=0}^{tM-1} I_{q,5}[n],~~
\frac{1}{tM}\sum_{n=0}^{tM-1} Q_{q,5}[n]
\right]_{q=1}^{5}.
\label{eq:c5}
\end{equation}
The $L_5$ and $C_5$ features are visualised in Fig.~\ref{fig:l5-c5-features}, using the \(I\) and \(Q\) components of \(q_5\) to compare the 5-step and 10-step features.
The $5$ vs $10$ timestep comparison shows that the additional timesteps resolve features that could be important for classifying the qubit state, particularly in the $I$ traces of the $L_5$ feature. 

\subsection{Network Architecture \& Training}

We use either just the local $L_5$ features or the combination of the $L_5$ and cumulative $C_5$ features as inputs to the SNNs.
The features are standardised using the training split mean and standard deviation.
The data is split into $432000$ training traces and $48000$ validation traces, with a separate $1120000$-trace test set used only for final evaluation.
Each split has an equal number of traces from each of the $32$ five-qubit states.
All models are trained to predict the five readout bits.
The class labels \(y_i\in\{0,\ldots,31\}\) are converted to a bit vector \(\mathbf{b}_i\in\{0,1\}^{5}\) for compatibility with the $5$-dim output layer used in the networks.

Each SNN has either a $10$-dim or a $20$-dim input layer.
The input layer is followed by a leaky integrate-and-fire (LIF) hidden layer with either $32$, $64$, or $128$ neurons and a dense $5$-dim output layer.
We also have an ANN baseline that uses a single hidden layer with $128$ neurons over the flattened time-series features.
The input dimension for the ANN is either $10$ for the $L_5$ features or $20$ for the $L_5+C_5$ features, multiplied by the number of chunks, either $5$ or $10$.
The input dimension for the ANN, therefore, ranges from $50$ to $200$.

The ANN baseline flattens the full sequence \(\mathbf{x}_i\in\mathbb{R}^{T\times d}\) and so the architecture can be summarised as
\[
\mathbb{R}^{Td}
\;\xrightarrow{\mathrm{Linear}}\;
128
\;\xrightarrow{\mathrm{ReLU}}\;
128
\;\xrightarrow{\mathrm{Dropout}(0.02)}\;
128
\;\xrightarrow{\mathrm{Linear}}\;
5 .
\]
The SNN processes one timestep at a time using a dense layer, one leaky integrate-and-fire hidden layer, dropout, and a dense five-bit readout, and so can be summarised as
\[
\mathbb{R}^{d}
\;\xrightarrow{\mathrm{Linear}}\;
h
\;\xrightarrow{\mathrm{LIF}}\;
h
\;\xrightarrow{\mathrm{Dropout}(0.02)}\;
h
\;\xrightarrow{\mathrm{Linear}}\;
5 ,
\qquad h\in\{32,64,128\}.
\]
The LIF hidden neurons use a learnable \(\beta\) parameter (initialised at \(\beta=0.75\)), a threshold of \(u_{\mathrm{th}}=1.0\), and a subtractive reset.  
The readout uses membrane accumulation with no readout decay \(\beta_{\mathrm{ro}}=1.0\), producing five logits \(\mathbf{z}_{i,t}\in\mathbb{R}^{5}\) after each timestep.
Final bit predictions are obtained by applying a fixed threshold at zero.
Fig.~\ref{fig:diagram-snn} depicts the chunk-by-chunk processing and the corresponding updates to the internal states and qubit assignments.

For a minibatch \(\mathcal{B}\), the training loss is bitwise binary cross-entropy with logits. 
Let
\[
\ell(\mathbf{z}_{i,t},\mathbf{b}_i)
=
\frac{1}{5}\sum_{q=1}^{5}
\left[
-b_{iq}\log\sigma(z_{iq,t})
-(1-b_{iq})\log\left(1-\sigma(z_{iq,t})\right)
\right].
\]
For the default paper models, the SNN loss is applied only after the final chunk,
\[
\mathcal{L}
=
\frac{1}{|\mathcal{B}|}
\sum_{i\in\mathcal{B}}
\ell(\mathbf{z}_{i,T},\mathbf{b}_i).
\]
The ANN uses the same bitwise binary cross-entropy objective for its single five-logit output.

Models are trained with a minibatch size of $1024$, and the best checkpoint is selected based on the validation geometric-mean assignment fidelity.  
Early stopping is triggered after $30$ epochs without an improvement larger than \(10^{-5}\).  
The maximum number of epochs allowed is $250$, but all models triggered early stopping before this.
All reported model results in the paper are obtained from a single training run with a fixed random seed, due to the large number of FP32 and QAT configurations studied and subsequently synthesised with \hlsfml{}.
However, repeated training runs indicate that the fidelity fluctuates at the level of $\mathcal{O}(10^{-4})$ with different seeds.
A summary of the hyperparameters used is given in Tab.~\ref{tab:training-hyperparameters}.

\begin{table}[t]
\centering
\small
\begin{tabular}{lll}
\toprule
Hyperparameter & ANN & SNN \\
\midrule
Hidden size(s) & \(128\) & \(32,64,128\) \\
Input features & \(L_5,\;L_5+C_5\) & \(L_5,\;L_5+C_5\) \\
Timesteps & \(5,10\) & \(5,10\) \\
Dropout & \(0.02\) & \(0.02\) \\
Optimiser & AdamW & AdamW \\
Learning rate & \(1.2\times10^{-3}\) & \(1.0\times10^{-2}\) \\
Weight decay & \(1.0\times10^{-4}\) & \(0\) \\
LR scheduler & ReduceLROnPlateau & ReduceLROnPlateau \\
Scheduler factor & \(0.5\) & \(0.5\) \\
Scheduler patience & 10 epochs & 10 epochs \\
Minimum learning rate & \(1.0\times10^{-5}\) & \(1.0\times10^{-5}\) \\
Maximum epochs & $250$ & $250$ \\
Early stopping patience & $30$ epochs & $30$ epochs \\
Selection metric & validation \(F_{\mathrm{geom}}\) & validation \(F_{\mathrm{geom}}\) \\
\bottomrule
\end{tabular}
\caption{Model and training hyperparameters used in training the ANN baseline and the SNN.}
\label{tab:training-hyperparameters}
\end{table}

\subsection{Results}

Tab.~\ref{tab:fp32-results} compares the MF, ANN, and representative SNN configurations.
We see that the 10-step ANN gives the highest score here, with \(F_{\mathrm{geom}}=0.9100\).
The best SNN result in this scan is the SNN with a $128$-dim hidden layer, the $L_5+C_5$ feature set, and $T\!=\!10$, which achieves \(F_{\mathrm{geom}}=0.9084\).
The $64$-dim \(L_5+C_5\) SNN is close, with \(F_{\mathrm{geom}}=0.9083\).
Because the ANN (i) has access to the full time-series of data in one forward pass and (ii) has a much larger network size, we expect its performance to be greater than that of the SNNs.
The key point here is that the performance of the ANN is not significantly better, which motivates the trade-off for early classification and real-time updates when using SNNs.

\begin{table}[t]
\centering
\small
\begin{tabular}{llrrrrrr}
\toprule
Model & Features & \(F_{\mathrm{geom}}\) & \(F_{q1}\) & \(F_{q2}\) & \(F_{q3}\) & \(F_{q4}\) & \(F_{q5}\) \\
\midrule
Matched filter & -- & 0.8962 & 0.9688 & 0.7396 & 0.9007 & 0.9360 & 0.9569 \\
ANN, $T\!=\!5$, $128$-dim hidden & \(L_5+C_5\) & 0.9087 & 0.9698 & 0.7418 & 0.9393 & 0.9459 & 0.9694 \\
ANN, $T\!=\!10$, $128$-dim hidden & \(L_5\) & 0.9100 & 0.9705 & 0.7435 & 0.9416 & 0.9465 & 0.9704 \\
SNN, $T\!=\!5$, $64$-dim hidden & \(L_5+C_5\) & 0.9070 & 0.9688 & 0.7404 & 0.9372 & 0.9449 & 0.9665 \\
SNN, $T\!=\!10$, $64$-dim hidden & \(L_5+C_5\) & 0.9083 & 0.9695 & 0.7421 & 0.9388 & 0.9453 & 0.9683 \\
SNN, $T\!=\!10$, $128$-dim hidden & \(L_5+C_5\) & 0.9084 & 0.9693 & 0.7424 & 0.9388 & 0.9454 & 0.9685 \\
\bottomrule
\end{tabular}
\caption{Summary of the readout results for the ANN baseline and the SNN models.}
\label{tab:fp32-results}
\end{table}

\begin{figure}[!htbp]
\centering
\includegraphics[width=0.62\linewidth]{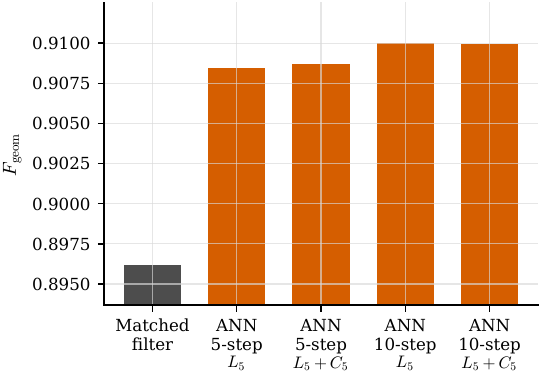}
\includegraphics[width=0.62\linewidth]{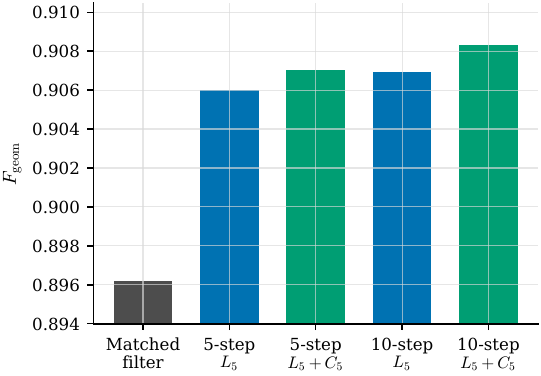}
\caption{
In the plot, we compare ANN models using the geometric mean assignment fidelity.
In the plot below, we compare SNN models with a $64$-dim hidden layer using the geometric mean assignment fidelity.}
\label{fig:fp32-models-snn-64-bars}
\end{figure}

We visualise the comparison between different hyperparameter choices for the ANN and SNN ($64$-dim) in Fig.~\ref{fig:fp32-models-snn-64-bars}.
The cumulative feature $C_5$ improves both the $T\!=\!5$ and $T\!=\!10$ models, with the $T\!=\!10$ \(L_5+C_5\) model giving the best SNN result.
However, this is not solely due to the information contained in $C_5$, as both the ANN and the SNN can reconstruct the cumulative features internally from just the $L_5$ features.
The driving factor here is simply that adding input features increases the model's parameter count, thereby increasing the network's capacity to perform its task.
This is particularly impactful for the SNN, where we have chosen network hyperparameters to meet the strict latency and resource requirements of the real-time readout.

\begin{figure}[!htbp]
\centering
\includegraphics[width=0.46\linewidth]{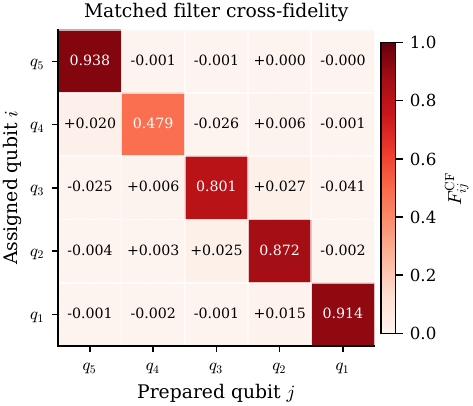}\hfill
\includegraphics[width=0.46\linewidth]{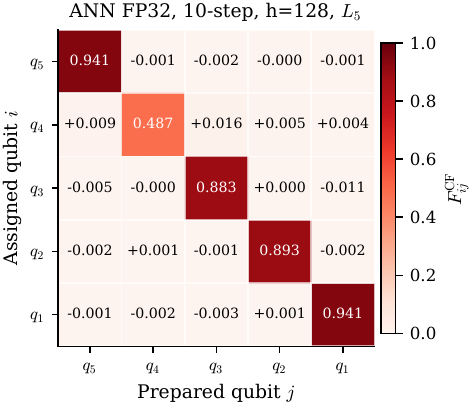}
\includegraphics[width=0.46\linewidth]{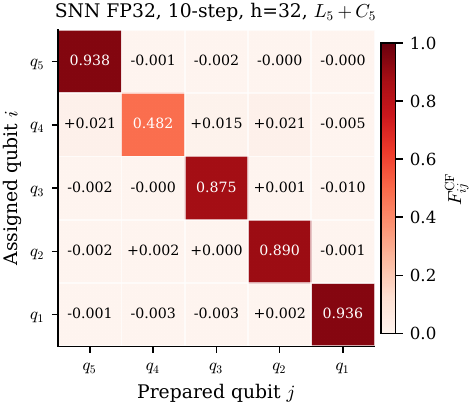}\hfill
\includegraphics[width=0.46\linewidth]{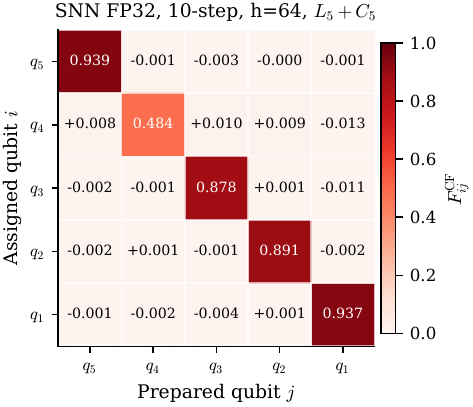}
\caption{Cross-fidelity matrices for MF baseline (top left), the ANN baseline (top right), the $32$-dim SNN (bottom left), and $64$-dim SNN (bottom right). All models use $T\!=\!10$ and the $L_5+C_5$ feature set.}
\label{fig:fp32-cross-talk}
\end{figure}

Fig.~\ref{fig:fp32-cross-talk} compares the cross-fidelity matrices for the ANN baseline and the $32$- and $64$-dim $T\!=\!10$ \(L_5+C_5\) SNNs.
All three learned classifiers reduce the off-diagonal structure seen in the MF benchmark.
The largest off-diagonal magnitude is \(0.0159\) for the ANN, \(0.0210\) for the $32$-dim SNN, and \(0.0130\) for the $64$-dim SNN.
The SNN with the $64$-dim hidden layer, therefore, gives the cleanest pairwise crosstalk matrix among these three representative models. At the same time, the ANN keeps a slightly larger diagonal contrast on most qubits, consistent with its marginally higher \(F_{\mathrm{geom}}\).
In all cases \(q_2\) remains the dominant limitation with the ML models improving its diagonal cross-fidelity from the MF value \(F^{\mathrm{CF}}_{22}=0.479\) only to about \(0.48\!-\!0.49\), so the gains for the ML models here mostly come from modest improvements across the other qubits.

\clearpage

\subsection{Mid-readout optimization}
\label{sec:early-classification}

The SNN produces an updated five-bit assignment after each readout chunk, so we can optimise the network on these intermediate results as well as the result at the end of the full \(1\,\mu\mathrm{s}\) window to target high-fidelity results at earlier times.
Optimising the SNN at different readout times like this makes it easier to find an optimal trade-off between decision latency and readout fidelity.
The baseline models above are trained with a loss on the final timestep only.
To test whether the same architecture can be biased toward earlier decisions, we also train the $64$-dim SNN on the \(L_5+C_5\) features with an additional binary-cross-entropy term at \(600\,\mathrm{ns}\), weighted equally with the final-time loss:
\[
\mathcal{L}_{600+\mathrm{final}}
=
\frac{1}{2|\mathcal{B}|}
\sum_{i\in\mathcal{B}}
\left[
\ell(\mathbf{z}_{i,t_{600}},\mathbf{b}_i)
+
\ell(\mathbf{z}_{i,T},\mathbf{b}_i)
\right],
\]
where \(t_{600}=3\) for the $T\!=\!5$ model and \(t_{600}=6\) for the $T\!=\!10$ model.  

\begin{figure}[t]
\centering
\includegraphics[width=0.90\linewidth]{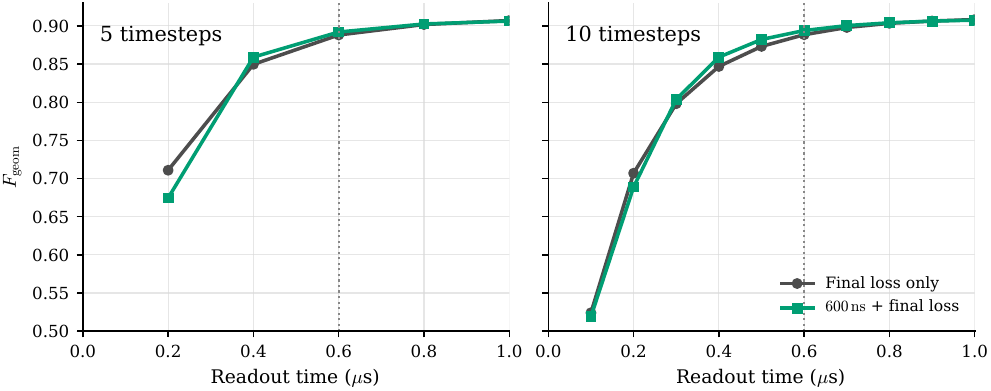}
\caption{Mid-readout fidelity for the $64$-dim SNN trained on \(L_5+C_5\) features. The final-loss-only model is compared with a model trained with equal loss terms at \(600\,\mathrm{ns}\) and \(1\,\mu\mathrm{s}\). The dotted vertical line marks the auxiliary-loss timestep.}
\label{fig:snn-early-loss600}
\end{figure}

\begin{figure}[t]
\centering
\includegraphics[width=0.50\linewidth]{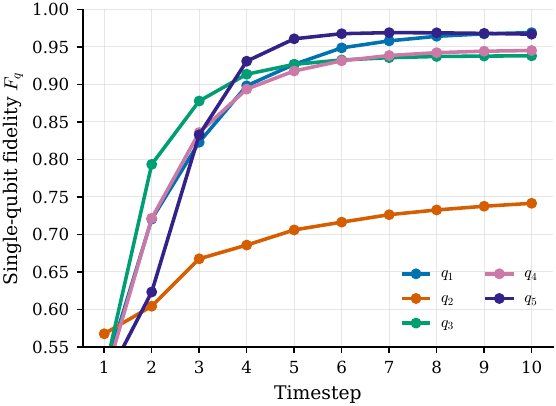}
\caption{Mid-readout fidelity for each qubit individually with the $64$-dim SNN.}
\label{fig:snn-early-loss600-perqubit}
\end{figure}

Fig.~\ref{fig:snn-early-loss600} and Tab.~\ref{tab:snn-loss600-results} compare the resulting classification fidelity as a function of readout time for the $T\!=\!5$ and $T\!=\!10$ models.
The auxiliary loss does not simply raise the whole curve.
It shifts performance toward the middle of the readout window: the earliest chunk is slightly worse, the \(300\!-\!800\,\mathrm{ns}\) region improves, and the final \(1\,\mu\mathrm{s}\) fidelity is essentially unchanged.
For the $T\!=\!5$ model, the \(600\,\mathrm{ns}\) fidelity increases from $0.8881$ to $0.8920$, while the final fidelity changes from $0.9070$ to $0.9063$.
For the $T\!=\!10$ model, the \(600\,\mathrm{ns}\) fidelity increases from $0.8885$ to $0.8939$, with the largest gain at \(400\,\mathrm{ns}\), and the final fidelity changes from $0.9083$ to $0.9076$.
This shows that a small change in the training objective can provide a more useful intermediate classifier without sacrificing the final readout assignment.
The value of the early classification is clearer when we look at the per-qubit fidelities produced by the SNN as a function of readout time.
Fig.~\ref{fig:snn-early-loss600-perqubit} shows that the rate at which the SNN gains confidence in the readout for each qubit is different.
We see that for $q_5$ and $q_3$ the readout fidelity has plateaued already by $600$ns, while $q_4$, $q_2$, and $q_1$ improve right up to the end of the window.

\begin{table}[t]
\centering
\small
\begin{tabular}{rrrrr}
\toprule
Timesteps & Time & Final loss only & \(600\,\mathrm{ns}\) + final loss & Change \\
\midrule
5 & \(600\,\mathrm{ns}\) & 0.8881 & 0.8920 & \(+0.0039\) \\
5 & \(1\,\mu\mathrm{s}\) & 0.9070 & 0.9063 & \(-0.0007\) \\
10 & \(600\,\mathrm{ns}\) & 0.8885 & 0.8939 & \(+0.0055\) \\
10 & \(1\,\mu\mathrm{s}\) & 0.9083 & 0.9076 & \(-0.0007\) \\
\bottomrule
\end{tabular}
\caption{Effect of the \(600\,\mathrm{ns}\) auxiliary loss on the $64$-dim \(L_5+C_5\) SNN.}
\label{tab:snn-loss600-results}
\end{table}

\section{FPGA Synthesis and Hardware Estimates}
\label{sec:fpga}

The shortest readout chunks considered in this work have a duration of $100\,\mathrm{ns}$.
For streaming readout, the central hardware requirement is that each SNN inference update completes within $100\,\mathrm{ns}$, so that the network state and assignment can be updated before the next chunk arrives.

The \hlsfml{} package has been used extensively in real-time inference for scientific applications, especially where models need to be designed with resource and latency constraints in mind.
The package has been extended to include convolutional layers~\cite{Aarrestad:2021zos,Ghielmetti:2022ndm}, distributed arithmetic~\cite{Sun:2025}, binary/ternary compatibility~\cite{Loncar:2020hqp}, and SNN inference~\cite{dillon2026spikingneuralnetworkinference}.

The results presented in this section are estimates provided by the AMD Vitis tool for High Level Synthesis from C-language descriptions.
Future work will explore the full demodulation, feature construction, data transfer, or FPGA place-and-route stages of a complete readout pipeline.

\subsection{Quantisation Aware Training}

All results thus far use full $32$-bit floating point numbers for all network weights and neuron dynamics.
Before synthesising the networks for FPGAs, we fine-tune the training to target the particular fixed-point precision that the networks will use on the FPGA.
This is called Quantisation-aware training (QAT).
For each SNN, we use \(W\)-bit fixed-point weights and neuron potentials with 4 integer bits, where \(W\in\{8,10,12,16\}\).
We fix the readout membrane state to use $16$ total bits, of which $8$ are integer bits.
Throughout the networks, we use rounding rather than truncation to bring numbers to the required precision, and we use saturation rather than wrapping when numbers fall outside the range of the bit precision.
The only exception is the accumulator, where we use truncation and wrapping.
This is for latency reasons, since adding rounding and saturation to the accumulator can significantly slow down the forward pass on the FPGA.
To counter this, we use $18$ total bits and $8$ integer bits in the accumulator.
The QAT routines that we use with the SNNs take these effects into account.

The QAT phase uses the AdamW optimiser with a learning rate \(5\times10^{-4}\), zero weight decay, a maximum of $100$ epochs, and early stopping with a patience of $20$ epochs.
It also uses a ReduceLROnPlateau scheduler with a factor \(0.5\), a patience of $8$ epochs, and a minimum learning rate \(10^{-5}\).  
The same bitwise loss and validation-fidelity model selection are used for QAT as for the previous training.
The results are shown in Tab.~\ref{tab:qat-results} where we see that the 10-step \(L_5+C_5\) model loses only \(5\times 10^{-4}\) in \(F_{\mathrm{geom}}\) at \(W=8\), and saturates by \(W=10\).
The local-only feature set is consistently lower than \(L_5+C_5\), partly because the larger input feature set yields a larger model, providing more capacity to learn features.
The results in all cases approach saturation at $W\!=\!12$.

\begin{figure}[t]
\centering
\includegraphics[width=0.58\linewidth]{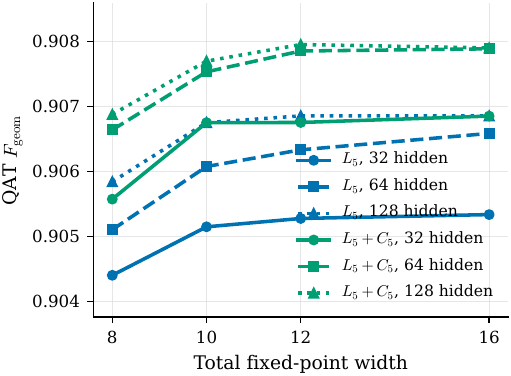}
\caption{Test fidelity results for the QAT precision sweep with SNNs with both $64$- and $32$-dim hidden layers, and chunk sizes of $100$ns ($T\!=\!10$).}
\label{fig:qat-precision}
\end{figure}

\begin{table}[t]
\centering
\small
\begin{tabular}{llrrrrrr}
\toprule
Features & Precision & \(F_{\mathrm{geom}}\) & \(F_{q1}\) & \(F_{q2}\) & \(F_{q3}\) & \(F_{q4}\) & \(F_{q5}\) \\
\midrule
\(L_5\) & \apfixeds{8}{4}  & 0.9051 & 0.9680 & 0.7383 & 0.9309 & 0.9442 & 0.9671 \\
\(L_5\) & \apfixeds{10}{4} & 0.9061 & 0.9683 & 0.7401 & 0.9327 & 0.9447 & 0.9672 \\
\(L_5\) & \apfixeds{12}{4} & 0.9063 & 0.9683 & 0.7411 & 0.9328 & 0.9446 & 0.9672 \\
\(L_5\) & \apfixeds{16}{4} & 0.9066 & 0.9684 & 0.7412 & 0.9337 & 0.9448 & 0.9673 \\
\(L_5+C_5\) & \apfixeds{8}{4}  & 0.9066 & 0.9688 & 0.7405 & 0.9370 & 0.9447 & 0.9646 \\
\(L_5+C_5\) & \apfixeds{10}{4} & 0.9075 & 0.9693 & 0.7414 & 0.9373 & 0.9449 & 0.9673 \\
\(L_5+C_5\) & \apfixeds{12}{4} & 0.9079 & 0.9694 & 0.7419 & 0.9378 & 0.9449 & 0.9676 \\
\(L_5+C_5\) & \apfixeds{16}{4} & 0.9079 & 0.9694 & 0.7420 & 0.9378 & 0.9449 & 0.9677 \\
\bottomrule
\end{tabular}
\caption{QAT results for the SNN with $T\!=\!10$ and a $64$-dim hidden layer.}
\label{tab:qat-results}
\end{table}

\subsection{Hardware Benchmarks}

The hardware benchmarks are obtained using the SNN functionality in \texttt{hls4ml}.
The synthesis sweep uses the 10-step SNN configurations selected for FPGA synthesis, covering hidden sizes \(h\in\{32,64,128\}\) and both the \(L_5\) and \(L_5+C_5\) feature sets.  
As summarised in Tab.~\ref{tab:hardware-hyperparams}, all models are generated with \hlsfml{} using the Vitis backend, a fully parallel I/O interface, latency-optimised dense layers with reuse factor one, and an \(8\,\mathrm{ns}\) target clock on the XCZU49DR FPGA from the Zynq UltraScale+ family (\texttt{xczu49dr-ffvf1760-2-e} device).  
The weight and activation precision is swept over \(\apfixed{W}{4}\) for \(W\in\{8,10,12,16\}\), while the readout layer uses \texttt{ap\_fixed<16,8>} and accumulators use \texttt{ap\_fixed<18,8>}.

\begin{table}
\centering
\small
\begin{tabular}{ll}
\toprule
Hyperparameter & Value \\
\midrule
Precision sweep & \({W}{4}\), \(W\in\{8,10,12,16\}\) \\
Readout precision & \texttt{<16,8>} \\
Accumulator precision & \texttt{<18,8>} \\
Backend & Vitis \\
FPGA part & \texttt{xczu49dr-ffvf1760-2-e} \\
I/O interface & \texttt{io\_parallel} \\
Target clock period & \(8.0\,\mathrm{ns}\) \\
Clock uncertainty & \(18\%\) \\
Configuration granularity & Layer name \\
Global reuse factor & \(1\) \\
Global strategy & Latency \\
\bottomrule
\end{tabular}
\caption{Hardware hyperparameters used for the FPGA synthesis sweep.}
\label{tab:hardware-hyperparams}
\end{table}

\begin{table}[t]
\centering
\scriptsize
\setlength{\tabcolsep}{3pt}
\resizebox{\linewidth}{!}{%
\begin{tabular}{llrrrrrrrrrrrr}
\toprule
Features & Hidden & Precision & \begin{tabular}{c}Clk est.\\(ns)\end{tabular} & II &
\begin{tabular}{c}Lat.\\cyc.\end{tabular} &
\begin{tabular}{c}Lat.\\(ns)\end{tabular} &
LUT & \begin{tabular}{c}LUT\\(\%)\end{tabular} &
FF & \begin{tabular}{c}FF\\(\%)\end{tabular} &
DSP & \begin{tabular}{c}DSP\\(\%)\end{tabular} \\
\midrule
$L_5$ & 32 & \(\langle 8,4\rangle\) & 6.4 & 1 & 5 & 32 & 22,793 & 5.4 & 1,736 & 0.2 & 0 & 0.0 \\
$L_5$ & 32 & \(\langle 10,4\rangle\) & 7.7 & 1 & 5 & 39 & 34,569 & 8.1 & 1,960 & 0.2 & 0 & 0.0 \\
$L_5$ & 32 & \(\langle 12,4\rangle\) & 7.7 & 1 & 4 & 31 & 29,251 & 6.9 & 1,846 & 0.2 & 247 & 5.8 \\
$L_5$ & 32 & \(\langle 16,4\rangle\) & 8.0 & 1 & 4 & 32 & 29,442 & 6.9 & 2,181 & 0.3 & 441 & 10.3 \\ \hdashline
$L_5$ & 64 & \(\langle 8,4\rangle\) & 6.5 & 1 & 6 & 39 & 44,601 & 10.5 & 3,142 & 0.4 & 0 & 0.0 \\
$L_5$ & 64 & \(\langle 10,4\rangle\) & 7.7 & 1 & 5 & 39 & 67,527 & 15.9 & 3,822 & 0.4 & 0 & 0.0 \\
$L_5$ & 64 & \(\langle 12,4\rangle\) & 7.7 & 1 & 4 & 31 & 57,547 & 13.5 & 3,599 & 0.4 & 502 & 11.8 \\
$L_5$ & 64 & \(\langle 16,4\rangle\) & 8.0 & 1 & 4 & 32 & 57,007 & 13.4 & 3,962 & 0.5 & 896 & 21.0 \\ \hdashline
$L_5$ & 128 & \(\langle 8,4\rangle\) & 6.6 & 1 & 8 & 52 & 85,572 & 20.1 & 7,462 & 0.9 & 0 & 0.0 \\
$L_5$ & 128 & \(\langle 10,4\rangle\) & 7.7 & 1 & 5 & 39 & 129,457 & 30.4 & 7,391 & 0.9 & 0 & 0.0 \\
$L_5$ & 128 & \(\langle 12,4\rangle\) & 7.7 & 1 & 4 & 31 & 112,476 & 26.4 & 6,857 & 0.8 & 1,000 & 23.4 \\
$L_5$ & 128 & \(\langle 16,4\rangle\) & 8.0 & 1 & 5 & 40 & 111,756 & 26.3 & 9,706 & 1.1 & 1,825 & 42.7 \\ \hdashline
$L_5+C_5$ & 32 & \(\langle 8,4\rangle\) & 6.5 & 1 & 5 & 33 & 30,675 & 7.2 & 2,164 & 0.3 & 0 & 0.0 \\
$L_5+C_5$ & 32 & \(\langle 10,4\rangle\) & 7.7 & 1 & 5 & 39 & 48,188 & 11.3 & 2,651 & 0.3 & 0 & 0.0 \\
$L_5+C_5$ & 32 & \(\langle 12,4\rangle\) & 7.7 & 1 & 4 & 31 & 38,911 & 9.1 & 2,757 & 0.3 & 403 & 9.4 \\
$L_5+C_5$ & 32 & \(\langle 16,4\rangle\) & 8.0 & 1 & 5 & 40 & 38,414 & 9.0 & 3,098 & 0.4 & 710 & 16.6 \\ \hdashline
$L_5+C_5$ & 64 & \(\langle 8,4\rangle\) & 6.5 & 1 & 6 & 39 & 52,276 & 12.3 & 3,952 & 0.5 & 0 & 0.0 \\
$L_5+C_5$ & 64 & \(\langle 10,4\rangle\) & 7.7 & 1 & 5 & 39 & 93,057 & 21.9 & 5,121 & 0.6 & 0 & 0.0 \\
$L_5+C_5$ & 64 & \(\langle 12,4\rangle\) & 7.7 & 1 & 4 & 31 & 75,760 & 17.8 & 5,105 & 0.6 & 817 & 19.1 \\
$L_5+C_5$ & 64 & \(\langle 16,4\rangle\) & 8.0 & 1 & 5 & 40 & 74,036 & 17.4 & 5,635 & 0.7 & 1,464 & 34.3 \\ \hdashline
$L_5+C_5$ & 128 & \(\langle 8,4\rangle\) & 6.5 & 1 & 8 & 52 & 102,757 & 24.2 & 8,977 & 1.1 & 0 & 0.0 \\
$L_5+C_5$ & 128 & \(\langle 10,4\rangle\) & 7.7 & 1 & 5 & 39 & 182,950 & 43.0 & 10,090 & 1.2 & 0 & 0.0 \\
$L_5+C_5$ & 128 & \(\langle 12,4\rangle\) & 7.7 & 1 & 4 & 31 & 149,219 & 35.1 & 9,767 & 1.1 & 1,718 & 40.2 \\
$L_5+C_5$ & 128 & \(\langle 16,4\rangle\) & 8.0 & 1 & 5 & 40 & 146,346 & 34.4 & 10,714 & 1.3 & 2,986 & 69.9 \\
\bottomrule
\end{tabular}
}
\caption{Per-timestep Vitis HLS C-synthesis estimates for the SNN hardware sweep on the XCZU49DR target.  
All runs use the parameters defined in Tab.~\ref{tab:hardware-hyperparams}.  
Latency is reported per HLS top-call/chunk.}
\label{tab:hardware-results}
\end{table}

The synthesis results in Tab.~\ref{tab:hardware-results} show that all configurations achieve an initiation interval of one, with a per-timestep latency between $31$ and $52\mathrm{ns}$.
This is below the $100\mathrm{ns}$ duration of the shortest readout chunks considered in this work.
The SNN inference can therefore be completed before the next chunk becomes available, allowing the network state and qubit assignments to be updated continuously during the acquisition.
The reported latency is for the SNN inference kernel only and does not include demodulation, averaging, or data transfer logic.

As expected, the resource usage increases with the hidden-layer size and with the number of input features.
Increasing the hidden size from $32$ to $128$ approximately scales the number of weights, and consequently the LUT and DSP requirements, by the same factor.
The $L_5+C_5$ models require more resources than the corresponding $L_5$ models because the input dimension is doubled, although the additional cumulative features improve the readout fidelity.
The FF usage remains below $1.3$\% for all configurations, so the dominant hardware trade-off is between LUT and DSP utilisation.

The choice of numerical precision also changes how the arithmetic is mapped to the device.
The 8- and 10-bit configurations use no DSP blocks, with multiplication implemented instead using LUT resources.
At 12 and 16 bits, the synthesis tool maps an increasing fraction of the arithmetic to DSP blocks, which can reduce the LUT usage and, in several configurations, the number of latency cycles.
The 16-bit models marginally exceed the $8$ns target clock estimate, reaching $8.026$ns, while the lower-precision models meet the target.
These C-synthesis estimates demonstrate that the SNN inference kernel can be executed within the interval between successive readout chunks on the target FPGA.

\section{Conclusions}
\label{sec:conclusions}

In this work, we have proposed and studied Spiking Neural Networks for real-time state assignment from frequency-multiplexed superconducting-qubit readout traces.
Unlike classifiers that operate only after the complete readout trace has been collected, the SNN processes successive chunks of the measurement record in real-time as the data arrives and maintains an internal temporal state.
This allows the SNN to start the state-assignment process earlier, maximising resource utilisation in a physically constrained setting.
The availability of time-dependent readouts also enables the controller to use them before the full measurement window has elapsed.

On the five-qubit dataset, the best full-precision SNN achieves a geometric-mean assignment fidelity of $F_{\mathrm{geom}}=0.9085$, compared with $0.8960$ for the matched-filter benchmark and $0.9100$ for the ANN baseline.
The SNN therefore recovers most of the performance of the larger ANN while using a compact architecture that processes only one readout chunk at a time.
The learned classifiers also reduce the off-diagonal structure in the cross-fidelity matrix.
For the representative $64$-dim SNN, the largest off-diagonal magnitude is reduced to $0.0130$, compared with $0.041$ for the MF, showing that joint processing of the multiplexed channels can suppress correlated readout errors.
Because the SNN produces an assignment after every timestep, its training objective can be modified to target high fidelity at different times during the readout.
This makes it easier to find an optimal trade-off between decision latency and readout fidelity.
Adding an auxiliary loss at $600$ns increases the $600$ns fidelity of the ten-step model from $0.8885$ to $0.8939$, while changing the final fidelity by only $7\times10^{-4}$.

QAT preserves the readout performance at the precision required for FPGA synthesis.
For the $64$-dim $L_5+C_5$ SNN, an 8-bit implementation reaches $F_{\mathrm{geom}}=0.9066$, while the performance is effectively saturated by 10-12 bit precision.
The FPGA synthesis results give per-timestep latencies of $31\!-\!52$ns, with an initiation interval of one for every configuration considered.
These latencies are below the $100$ns duration of the shortest input chunks, demonstrating that the SNN can update its internal state and prediction before the next chunk of measurement data arrives.
Several configurations require no DSP resources, providing additional flexibility when integrating the discriminator alongside the control and signal-processing logic of a quantum-computing system.
The latency and resource usage reported here will improve as the SNN functionality in \hlsfml is further developed.

The results establish SNNs as a promising approach to low-latency readout of superconducting qubits, combining temporal classification and hardware-efficient inference in a single model.
Further improvements could be achieved through confidence-based stopping criteria, timestep-dependent training objectives, and improved choices of spiking neurons.
Scaling the method to larger multiplexed systems will also determine how effectively the SNN can exploit temporal and inter-channel correlations as the number of simultaneously measured qubits increases.

\section*{Acknowledgment}
\label{sec:acknowledgment}
P.D. and B.L. acknowledge support from the German Federal Ministry of Research, Technology and Space (BMFTR) through the program EQuIPS (Grant No. 13N17232). 
B.D., J.H., and A.J. are grateful for the use of the computing resources from the Northern Ireland High Performance Computing (NI-HPC) service funded by EPSRC (EP/T022175).

\appendix
\section{Readout traces}
\counterwithin{figure}{section}
\setcounter{figure}{0}

\begin{figure}[!htbp]
\centering
\includegraphics[width=0.78\linewidth]{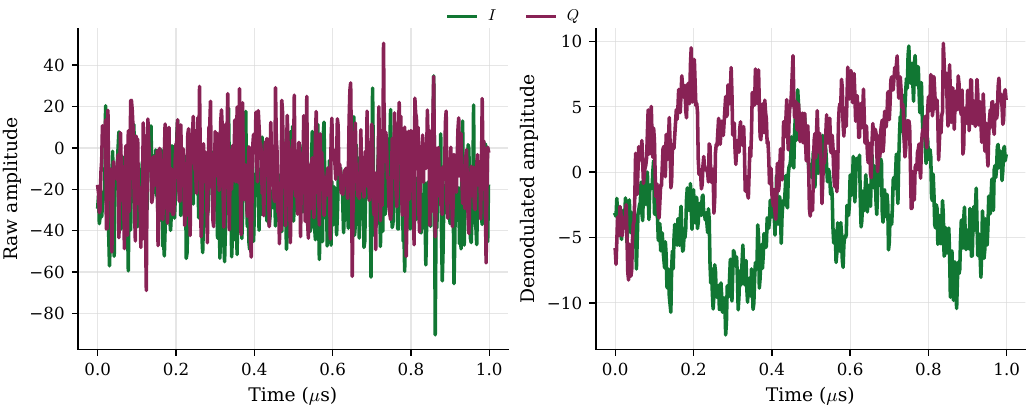}
\caption{Representative digitally-unmodulated and demodulated readout trace.}
\label{fig:readout-example}
\end{figure}

\begin{figure}[!htbp]
\centering
\includegraphics[width=0.85\linewidth]{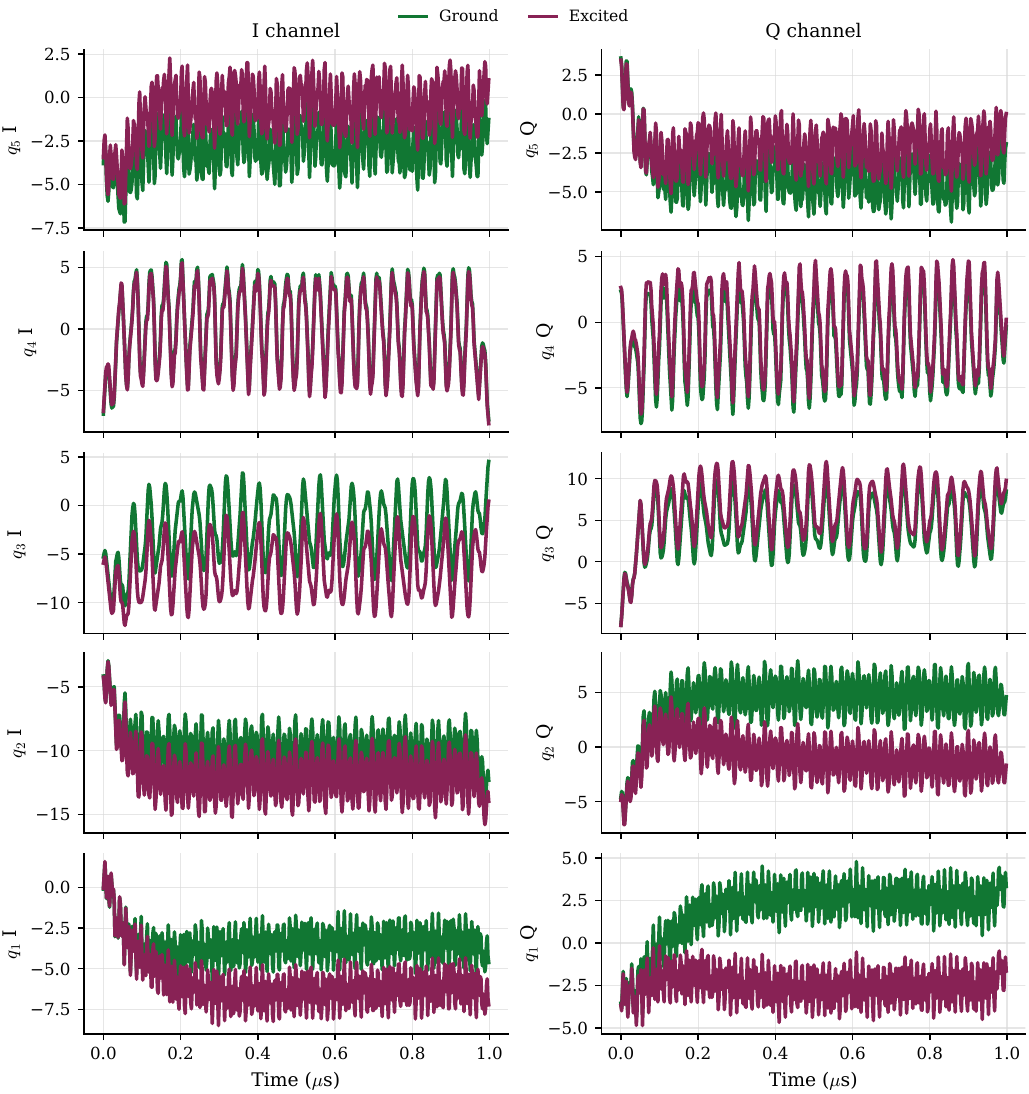}
\caption{Average demodulated ground- and excited-state traces for the five qubits.}
\label{fig:readout-average}
\end{figure}

In Fig.~\ref{fig:readout-example} and Fig.~\ref{fig:readout-average}, we show the trace for the first qubit and the average demodulated traces for all five qubits, respectively.

\bibliographystyle{unsrt}
\bibliography{refs}

@article{ANDERSON1972197,
title = {A simple neural network generating an interactive memory},
journal = {Mathematical Biosciences},
volume = {14},
number = {3},
pages = {197-220},
year = {1972},
issn = {0025-5564},
doi = {https://doi.org/10.1016/0025-5564(72)90075-2},
url = {https://www.sciencedirect.com/science/article/pii/0025556472900752},
author = {James A. Anderson}
}

@article{TAVANAEI201947,
title = {Deep learning in spiking neural networks},
journal = {Neural Networks},
volume = {111},
pages = {47-63},
year = {2019},
issn = {0893-6080},
doi = {https://doi.org/10.1016/j.neunet.2018.12.002},
url = {https://www.sciencedirect.com/science/article/pii/S0893608018303332},
author = {Amirhossein Tavanaei and Masoud Ghodrati and Saeed Reza Kheradpisheh and Timothée Masquelier and Anthony Maida},
}

@misc{oconnor2016deepspikingnetworks,
      title={Deep Spiking Networks},
      author={Peter O'Connor and Max Welling},
      year={2016},
      eprint={1602.08323},
      archivePrefix={arXiv},
      primaryClass={cs.NE},
      url={https://arxiv.org/abs/1602.08323},
}

@ARTICLE{8891809,
  author={Neftci, Emre O. and Mostafa, Hesham and Zenke, Friedemann},
  journal={IEEE Signal Processing Magazine},
  title={Surrogate Gradient Learning in Spiking Neural Networks: Bringing the Power of Gradient-Based Optimization to Spiking Neural Networks},
  year={2019},
  volume={36},
  number={6},
  pages={51-63},
  doi={10.1109/MSP.2019.2931595}
}

@inproceedings{10.1145/3183584.3183612,
author = {Schuman, Catherine D. and Potok, Thomas E. and Young, Steven and Patton, Robert and Perdue, Gabriel and Chakma, Gangotree and Wyer, Austin and Rose, Garrett S.},
title = {Neuromorphic computing for temporal scientific data classification},
year = {2017},
isbn = {9781450364423},
publisher = {Association for Computing Machinery},
address = {New York, NY, USA},
url = {https://doi.org/10.1145/3183584.3183612},
doi = {10.1145/3183584.3183612},
booktitle = {Proceedings of the Neuromorphic Computing Symposium},
articleno = {2},
numpages = {6},
location = {Knoxville, Tennessee},
series = {NCS '17}
}

@misc{kulkarni2023onsensordatafilteringusing,
      title={On-Sensor Data Filtering using Neuromorphic Computing for High Energy Physics Experiments},
      author={Shruti R. Kulkarni et al},
      _author={Shruti R. Kulkarni and Aaron Young and Prasanna Date and Narasinga Rao Miniskar and Jeffrey S. Vetter and Farah Fahim and Benjamin Parpillon and Jennet Dickinson and Nhan Tran and Jieun Yoo and Corrinne Mills and Morris Swartz and Petar Maksimovic and Catherine D. Schuman and Alice Bean},
      year={2023},
      eprint={2307.11242},
      archivePrefix={arXiv},
      primaryClass={cs.NE},
      url={https://arxiv.org/abs/2307.11242},
}

@article{Coradin_2025,
   title={Unsupervised Particle Tracking with Neuromorphic Computing},
   volume={8},
   ISSN={2571-712X},
   url={http://dx.doi.org/10.3390/particles8020040},
   DOI={10.3390/particles8020040},
   number={2},
   journal={Particles},
   publisher={MDPI AG},
   author={Coradin, Emanuele and Cufino, Fabio and Awais, Muhammad and Dorigo, Tommaso and Lupi, Enrico and Porcu, Eleonora and Raj, Jinu and Sandin, Fredrik and Tosi, Mia},
   year={2025},
   month=Apr, pages={40} }

@misc{lupi2025neuromorphicreadouthadroncalorimeters,
      title={Neuromorphic Readout for Hadron Calorimeters},
      author={Enrico Lupi et al},
      _author={Enrico Lupi and Abhishek and Max Aehle and Muhammad Awais and Alessandro Breccia and Riccardo Carroccio and Long Chen and Abhijit Das and Andrea De Vita and Tommaso Dorigo and Nicolas R. Gauger and Ralf Keidel and Jan Kieseler and Anders Mikkelsen and Federico Nardi and Xuan Tung Nguyen and Fredrik Sandin and Kylian Schmidt and Pietro Vischia and Joseph Willmore},
      year={2025},
      eprint={2502.12693},
      archivePrefix={arXiv},
      primaryClass={hep-ex},
      url={https://arxiv.org/abs/2502.12693},
}

@article{Dillon:2025yqe,
    author = "Dillon, Barry M. and Harkin, Jim and Javed, Aqib",
    title = "{Anomaly detection with spiking neural networks for LHC physics}",
    eprint = "2508.00063",
    archivePrefix = "arXiv",
    primaryClass = "hep-ph",
    doi = "10.1088/2632-2153/ae4a85",
    journal = "Mach. Learn. Sci. Tech.",
    volume = "7",
    number = "2",
    pages = "025019",
    year = "2026"
}

@misc{mehonic2024roadmapneuromorphiccomputingemerging,
      title={Roadmap to Neuromorphic Computing with Emerging Technologies},
      author={Adnan Mehonic et al},
      _author={Adnan Mehonic and Daniele Ielmini and Kaushik Roy and Onur Mutlu and Shahar Kvatinsky and Teresa Serrano-Gotarredona and Bernabe Linares-Barranco and Sabina Spiga and Sergey Savelev and Alexander G Balanov and Nitin Chawla and Giuseppe Desoli and Gerardo Malavena and Christian Monzio Compagnoni and Zhongrui Wang and J Joshua Yang and Ghazi Sarwat Syed and Abu Sebastian and Thomas Mikolajick and Beatriz Noheda and Stefan Slesazeck and Bernard Dieny and Tuo-Hung and Hou and Akhil Varri and Frank Bruckerhoff-Pluckelmann and Wolfram Pernice and Xixiang Zhang and Sebastian Pazos and Mario Lanza and Stefan Wiefels and Regina Dittmann and Wing H Ng and Mark Buckwell and Horatio RJ Cox and Daniel J Mannion and Anthony J Kenyon and Yingming Lu and Yuchao Yang and Damien Querlioz and Louis Hutin and Elisa Vianello and Sayeed Shafayet Chowdhury and Piergiulio Mannocci and Yimao Cai and Zhong Sun and Giacomo Pedretti and John Paul Strachan and Dmitri Strukov and Manuel Le Gallo and Stefano Ambrogio and Ilia Valov and Rainer Waser},
      year={2024},
      eprint={2407.02353},
      archivePrefix={arXiv},
      primaryClass={eess.SP},
      url={https://arxiv.org/abs/2407.02353},
}

@INPROCEEDINGS{8683736,
  author={Song, Linghao and Chen, Fan and Young, Steven R. and Schuman, Catherine D. and Perdue, Gabriel and Potok, Thomas E.},
  booktitle={ICASSP 2019 - 2019 IEEE International Conference on Acoustics, Speech and Signal Processing (ICASSP)},
  title={Deep Learning for Vertex Reconstruction of Neutrino-nucleus Interaction Events with Combined Energy and Time Data},
  year={2019},
  volume={},
  number={},
  pages={3882-3886},
  doi={10.1109/ICASSP.2019.8683736}}

@ARTICLE{10.3389/fphy.2024.1334298,
AUTHOR={Lin, S.  and Ning, S.  and Zhu, H.  and Zhou, T.  and Morris, C. L.  and Clayton, S.  and Cherukara, M. J.  and Chen, R. T.  and Wang, Zhehui },
TITLE={Neural network methods for radiation detectors and imaging},
JOURNAL={Frontiers in Physics},
VOLUME={Volume 12 - 2024},
YEAR={2024},
URL={https://www.frontiersin.org/journals/physics/articles/10.3389/fphy.2024.1334298},
DOI={10.3389/fphy.2024.1334298},
ISSN={2296-424X}
}

@article{Maass:1997spiking,
  author  = {Maass, Wolfgang},
  title   = {Networks of Spiking Neurons: The Third Generation of Neural Network Models},
  journal = {Neural Networks},
  volume  = {10},
  number  = {9},
  pages   = {1659--1671},
  year    = {1997},
  doi     = {10.1016/S0893-6080(97)00011-7}
}

@article{Merolla:2014million,
  author  = {Merolla, Paul A. and Arthur, John V. and Alvarez-Icaza, Rodrigo and Cassidy, Andrew S. and Sawada, Jun and Akopyan, Filipp and Jackson, Bryan L. and Imam, Nabil and Guo, Chen and Nakamura, Yutaka and others},
  title   = {A Million Spiking-Neuron Integrated Circuit with a Scalable Communication Network and Interface},
  journal = {Science},
  volume  = {345},
  number  = {6197},
  pages   = {668--673},
  year    = {2014},
  doi     = {10.1126/science.1254642}
}

@article{Davies:2018loihi,
  author  = {Davies, Mike and Srinivasa, Narayan and Lin, Tsung-Han and Chinya, Gautham and Cao, Yongqiang and Choday, Sri Harsha and Dimou, Georgios and Joshi, Prasad and Imam, Nabil and Jain, Shweta and others},
  title   = {Loihi: A Neuromorphic Manycore Processor with On-Chip Learning},
  journal = {IEEE Micro},
  volume  = {38},
  number  = {1},
  pages   = {82--99},
  year    = {2018},
  doi     = {10.1109/MM.2018.112130359}
}

@misc{shrestha2023efficientvideoaudioprocessing,
      title={Efficient Video and Audio processing with Loihi 2}, 
      author={Sumit Bam Shrestha and Jonathan Timcheck and Paxon Frady and Leobardo Campos-Macias and Mike Davies},
      year={2023},
      eprint={2310.03251},
      archivePrefix={arXiv},
      primaryClass={cs.NE},
      url={https://arxiv.org/abs/2310.03251}, 
}

@article{Furber:2014spinnaker,
  author  = {Furber, Steve B. and Galluppi, Francesco and Temple, Steve and Plana, Luis A.},
  title   = {The SpiNNaker Project},
  journal = {Proceedings of the IEEE},
  volume  = {102},
  number  = {5},
  pages   = {652--665},
  year    = {2014},
  doi     = {10.1109/JPROC.2014.2304638}
}

@article{Zenke_2018,
   title={SuperSpike: Supervised Learning in Multilayer Spiking Neural Networks},
   volume={30},
   ISSN={1530-888X},
   url={http://dx.doi.org/10.1162/neco_a_01086},
   DOI={10.1162/neco_a_01086},
   number={6},
   journal={Neural Computation},
   publisher={MIT Press - Journals},
   author={Zenke, Friedemann and Ganguli, Surya},
   year={2018},
   month={June},
   pages={1514–1541} 
}

@article{Eshraghian:2023lessons,
  author        = {Eshraghian, Jason K. and Ward, Max and Neftci, Emre O. and Wang, Xinxin and Lenz, Gregor and Dwivedi, Girish and Bennamoun, Mohammed and Jeong, Doo Seok and Lu, Wei D.},
  title         = {Training Spiking Neural Networks Using Lessons From Deep Learning},
  journal       = {Proceedings of the IEEE},
  volume        = {111},
  number        = {9},
  pages         = {1016--1054},
  year          = {2023},
  doi           = {10.1109/JPROC.2023.3308088},
  eprint        = {2109.12894},
  archivePrefix = {arXiv},
  primaryClass  = {cs.NE}
}

@misc{paszke2019pytorchimperativestylehighperformance,
      title={PyTorch: An Imperative Style, High-Performance Deep Learning Library},
      author={Adam Paszke et al},
      _author={Adam Paszke and Sam Gross and Francisco Massa and Adam Lerer and James Bradbury and Gregory Chanan and Trevor Killeen and Zeming Lin and Natalia Gimelshein and Luca Antiga and Alban Desmaison and Andreas Köpf and Edward Yang and Zach DeVito and Martin Raison and Alykhan Tejani and Sasank Chilamkurthy and Benoit Steiner and Lu Fang and Junjie Bai and Soumith Chintala},
      year={2019},
      eprint={1912.01703},
      archivePrefix={arXiv},
      primaryClass={cs.LG},
      url={https://arxiv.org/abs/1912.01703},
}

@article{Duarte:2018ite,
  author        = {Duarte, Javier and others},
  title         = {Fast Inference of Deep Neural Networks in {FPGAs} for Particle Physics},
  journal       = {Journal of Instrumentation},
  volume        = {13},
  number        = {07},
  pages         = {P07027},
  year          = {2018},
  doi           = {10.1088/1748-0221/13/07/P07027},
  eprint        = {1804.06913},
  archivePrefix = {arXiv},
  primaryClass  = {physics.ins-det},
  reportNumber  = {FERMILAB-PUB-18-089-E}
}

@inproceedings{Fahim:2021hls4mlCodesign,
  author        = {Fahim, Farah and Hawks, Benjamin and Herwig, Christian and Hirschauer, James and Jindariani, Sergo and Tran, Nhan and Carloni, Luca P. and Di Guglielmo, Giuseppe and Harris, Philip and Krupa, Jeffrey and others},
  title         = {{hls4ml}: An Open-Source Codesign Workflow to Empower Scientific Low-Power Machine Learning Devices},
  booktitle     = {Proceedings of the tinyML Research Symposium},
  year          = {2021},
  eprint        = {2103.05579},
  archivePrefix = {arXiv},
  primaryClass  = {cs.LG},
  url           = {https://openreview.net/forum?id=YBFqldo30vG}
}

@article{Schulte:2025hls4mlPlatform,
  author        = {Schulte, Jan-Frederik and others},
  title         = {{hls4ml}: A Flexible, Open-Source Platform for Deep Learning Acceleration on Reconfigurable Hardware},
  journal       = {ACM Transactions on Reconfigurable Technology and Systems},
  year          = {2026},
  note          = {Just accepted},
  doi           = {10.1145/3801979},
  eprint        = {2512.01463},
  archivePrefix = {arXiv},
  primaryClass  = {cs.AR}
}

@article{Aarrestad:2021zos,
  author        = {Aarrestad, Thea and others},
  title         = {Fast Convolutional Neural Networks on {FPGAs} with {hls4ml}},
  journal       = {Machine Learning: Science and Technology},
  volume        = {2},
  number        = {4},
  pages         = {045015},
  year          = {2021},
  doi           = {10.1088/2632-2153/ac0ea1},
  eprint        = {2101.05108},
  archivePrefix = {arXiv},
  primaryClass  = {cs.LG},
  reportNumber  = {FERMILAB-PUB-21-130-SCD}
}

@article{Ghielmetti:2022ndm,
    author = "Ghielmetti, Nicol\`{o} and others",
    title = "{Real-time semantic segmentation on FPGAs for autonomous vehicles with hls4ml}",
    eprint = "2205.07690",
    archivePrefix = "arXiv",
    primaryClass = "cs.CV",
    reportNumber = "FERMILAB-PUB-22-435-PPD",
    doi = "10.1088/2632-2153/ac9cb5",
    journal ="Mach. Learn. Sci. Tech.",
    year = "2022"
}

@misc{Sun:2025,
      title={da4ml: Distributed Arithmetic for Real-time Neural Networks on FPGAs},
      author={Chang Sun and others},
      year={2025},
      eprint={2507.04535},
      archivePrefix={arXiv},
      primaryClass={cs.AR},
      url={https://arxiv.org/abs/2507.04535},
}

@article{Loncar:2020hqp,
    author = "Ngadiuba, Jennifer and others",
    title = "{Compressing deep neural networks on FPGAs to binary and ternary precision with HLS4ML}",
    eprint = "2003.06308",
    archivePrefix = "arXiv",
    primaryClass = "cs.LG",
    reportNumber = "FERMILAB-PUB-20-167-PPD-SCD",
    doi = "10.1088/2632-2153/aba042",
    journal = "Mach. Learn. Sci. Tech.",
    volume = "2",
    pages = "015001",
    year = "2021"
}

@misc{dillon2026spikingneuralnetworkinference,
      title={Spiking Neural Network inference on FPGAs with hls4ml},
      author={Barry M. Dillon},
      year={2026},
      eprint={2606.10008},
      archivePrefix={arXiv},
      primaryClass={cs.NE},
      url={https://arxiv.org/abs/2606.10008},
}

@article{Magesan:2014bto,
    author = "Magesan, Easwar and Gambetta, Jay M. and C{\'o}rcoles, A. {\,}D. and Chow, Jerry M.",
    title = "{Machine Learning for Discriminating Quantum Measurement Trajectories and Improving Readout}",
    eprint = "1411.4994",
    archivePrefix = "arXiv",
    primaryClass = "quant-ph",
    doi = "10.1103/PhysRevLett.114.200501",
    journal = "Phys. Rev. Lett.",
    volume = "114",
    number = "20",
    pages = "200501",
    year = "2015"
}

@article{Flurin:2018rzr,
    author = "Flurin, Emmanuel and Martin, Leigh S. and Hacohen-Gourgy, Shay and Siddiqi, Irfan",
    title = "{Using a Recurrent Neural Network to Reconstruct Quantum Dynamics of a Superconducting Qubit from Physical Observations}",
    eprint = "1811.12420",
    archivePrefix = "arXiv",
    primaryClass = "quant-ph",
    doi = "10.1103/PhysRevX.10.011006",
    journal = "Phys. Rev. X",
    volume = "10",
    number = "1",
    pages = "011006",
    year = "2020"
}

@article{Martinez:2020bbx,
    author = "Martinez, Luis A. and Rosen, Yaniv J. and DuBois, Jonathan L.",
    title = "{Improving Qubit Readout with Hidden Markov Models}",
    eprint = "2006.00109",
    archivePrefix = "arXiv",
    primaryClass = "quant-ph",
    doi = "10.1103/PhysRevA.102.062426",
    journal = "Phys. Rev. A",
    volume = "102",
    pages = "062426",
    year = "2020"
}

@article{Lienhard:2022tiq,
    author = "Lienhard, Benjamin and others",
    title = "{Deep-Neural-Network Discrimination of Multiplexed Superconducting-Qubit States}",
    doi = "10.1103/PhysRevApplied.17.014024",
    journal = "Phys. Rev. Applied",
    volume = "17",
    number = "1",
    pages = "014024",
    year = "2022"
}

@article{Angelatos:2020fvv,
    author = {Angelatos, Gerasimos and Khan, Saeed and T{\"u}reci, Hakan E.},
    title = "{Reservoir Computing Approach to Quantum State Measurement}",
    eprint = "2011.09652",
    archivePrefix = "arXiv",
    primaryClass = "quant-ph",
    doi = "10.1103/PhysRevX.11.041062",
    journal = "Phys. Rev. X",
    volume = "11",
    pages = "041062",
    year = "2021"
}

@article{Kim:2021ixr,
    author = "Kim, Jihye and Oh, Byungdu and Chong, Yonuk and Hwang, Euyheon and Park, Daniel K.",
    title = "{Quantum readout error mitigation via deep learning}",
    eprint = "2112.03585",
    archivePrefix = "arXiv",
    primaryClass = "quant-ph",
    doi = "10.1088/1367-2630/ac7b3d",
    journal = "New J. Phys.",
    volume = "24",
    number = "7",
    pages = "073009",
    year = "2022"
}

@article{Reuer:2022efh,
    author = "Reuer, Kevin and others",
    title = "{Realizing a deep reinforcement learning agent for real-time quantum feedback}",
    eprint = "2210.16715",
    archivePrefix = "arXiv",
    primaryClass = "quant-ph",
    doi = "10.1038/s41467-023-42901-3",
    journal = "Nature Commun.",
    volume = "14",
    number = "1",
    pages = "7138",
    year = "2023"
}

@inproceedings{Maurya:2022ida,
    author = "Maurya, Satvik and Mude, Chaithanya Naik and Oliver, William D. and Lienhard, Benjamin and Tannu, Swamit",
    title = "{Scaling Qubit Readout with Hardware Efficient Machine Learning Architectures}",
    booktitle = "{50th Annual International Symposium on Computer Architecture}",
    eprint = "2212.03895",
    archivePrefix = "arXiv",
    primaryClass = "quant-ph",
    doi = "10.1145/3579371.3589042",
    month = "12",
    year = "2022"
}

@article{Cosco:2023jax,
    author = "Cosco, F. and Gullo, N. Lo",
    title = "{Enhancing qubit readout with Bayesian learning}",
    eprint = "2302.07725",
    archivePrefix = "arXiv",
    primaryClass = "quant-ph",
    doi = "10.1103/PhysRevA.108.L060402",
    journal = "Phys. Rev. A",
    volume = "108",
    number = "6",
    pages = "L060402",
    year = "2023"
}

@article{Barrett:2023nhg,
    author = "Barrett, Cora N. and others",
    title = "{Learning-Based Calibration of Flux Crosstalk in Transmon Qubit Arrays}",
    eprint = "2303.03347",
    archivePrefix = "arXiv",
    primaryClass = "quant-ph",
    doi = "10.1103/PhysRevApplied.20.024070",
    journal = "Phys. Rev. Applied",
    volume = "20",
    number = "2",
    pages = "024070",
    year = "2023"
}

@article{DiGuglielmo:2025zod,
    author = "Di Guglielmo, Giuseppe and others",
    title = "{End-to-End Workflow for Machine-Learning-Based Qubit Readout With QICK and hls4ml}",
    eprint = "2501.14663",
    archivePrefix = "arXiv",
    primaryClass = "quant-ph",
    reportNumber = "FERMILAB-PUB-24-0925-ETD-PPD",
    doi = "10.1109/TQE.2025.3604712",
    journal = "IEEE Trans. Quantum Eng.",
    volume = "6",
    pages = "1--10",
    year = "2025"
}

@article{Kent:2025fxy,
    author = "Kent, Robert and Lienhard, Benjamin and Lafyatis, Gregory and Gauthier, Daniel J.",
    title = "{Superconducting-qubit readout using next-generation reservoir computing}",
    eprint = "2506.15771",
    archivePrefix = "arXiv",
    primaryClass = "quant-ph",
    doi = "10.1103/bnwn-d2p4",
    journal = "Phys. Rev. Applied",
    volume = "25",
    number = "4",
    pages = "044009",
    year = "2026"
}

@article{10.1063/1.5089550,
    author = {Krantz, P. and Kjaergaard, M. and Yan, F. and Orlando, T. P. and Gustavsson, S. and Oliver, W. D.},
    title = {A quantum engineer's guide to superconducting qubits},
    journal = {Applied Physics Reviews},
    volume = {6},
    number = {2},
    pages = {021318},
    year = {2019},
    month = {06},
    issn = {1931-9401},
    doi = {10.1063/1.5089550},
    url = {https://doi.org/10.1063/1.5089550},
    eprint = {https://pubs.aip.org/aip/apr/article-pdf/doi/10.1063/1.5089550/20722375/021318_1_1.5089550.pdf},
}

@Article{Acharya2025,
author={AI, Google Quantum
and {Collaborators}},
_author={Acharya, Rajeev
and Abanin, Dmitry A.
and Aghababaie-Beni, Laleh
and Aleiner, Igor
and Andersen, Trond I.
and Ansmann, Markus
and Arute, Frank
and Arya, Kunal
and Asfaw, Abraham
and Astrakhantsev, Nikita
and Atalaya, Juan
and Babbush, Ryan
and Bacon, Dave
and Ballard, Brian
and Bardin, Joseph C.
and Bausch, Johannes
and Bengtsson, Andreas
and Bilmes, Alexander
and Blackwell, Sam
and Boixo, Sergio
and Bortoli, Gina
and Bourassa, Alexandre
and Bovaird, Jenna
and Brill, Leon
and Broughton, Michael
and Browne, David A.
and Buchea, Brett
and Buckley, Bob B.
and Buell, David A.
and Burger, Tim
and Burkett, Brian
and Bushnell, Nicholas
and Cabrera, Anthony
and Campero, Juan
and Chang, Hung-Shen
and Chen, Yu
and Chen, Zijun
and Chiaro, Ben
and Chik, Desmond
and Chou, Charina
and Claes, Jahan
and Cleland, Agnetta Y.
and Cogan, Josh
and Collins, Roberto
and Conner, Paul
and Courtney, William
and Crook, Alexander L.
and Curtin, Ben
and Das, Sayan
and Davies, Alex
and De Lorenzo, Laura
and Debroy, Dripto M.
and Demura, Sean
and Devoret, Michel
and Di Paolo, Agustin
and Donohoe, Paul
and Drozdov, Ilya
and Dunsworth, Andrew
and Earle, Clint
and Edlich, Thomas
and Eickbusch, Alec
and Elbag, Aviv Moshe
and Elzouka, Mahmoud
and Erickson, Catherine
and Faoro, Lara
and Farhi, Edward
and Ferreira, Vinicius S.
and Burgos, Leslie Flores
and Forati, Ebrahim
and Fowler, Austin G.
and Foxen, Brooks
and Ganjam, Suhas
and Garcia, Gonzalo
and Gasca, Robert
and Genois, {\'E}lie
and Giang, William
and Gidney, Craig
and Gilboa, Dar
and Gosula, Raja
and Dau, Alejandro Grajales
and Graumann, Dietrich
and Greene, Alex
and Gross, Jonathan A.
and Habegger, Steve
and Hall, John
and Hamilton, Michael C.
and Hansen, Monica
and Harrigan, Matthew P.
and Harrington, Sean D.
and Heras, Francisco J. H.
and Heslin, Stephen
and Heu, Paula
and Higgott, Oscar
and Hill, Gordon
and Hilton, Jeremy
and Holland, George
and Hong, Sabrina
and Huang, Hsin-Yuan
and Huff, Ashley
and Huggins, William J.
and Ioffe, Lev B.
and Isakov, Sergei V.
and Iveland, Justin
and Jeffrey, Evan
and Jiang, Zhang
and Jones, Cody
and Jordan, Stephen
and Joshi, Chaitali
and Juhas, Pavol
and Kafri, Dvir
and Kang, Hui
and Karamlou, Amir H.
and Kechedzhi, Kostyantyn
and Kelly, Julian
and Khaire, Trupti
and Khattar, Tanuj
and Khezri, Mostafa
and Kim, Seon
and Klimov, Paul V.
and Klots, Andrey R.
and Kobrin, Bryce
and Kohli, Pushmeet
and Korotkov, Alexander N.
and Kostritsa, Fedor
and Kothari, Robin
and Kozlovskii, Borislav
and Kreikebaum, John Mark
and Kurilovich, Vladislav D.
and Lacroix, Nathan
and Landhuis, David
and Lange-Dei, Tiano
and Langley, Brandon W.
and Laptev, Pavel
and Lau, Kim-Ming
and Le Guevel, Lo{\"i}ck
and Ledford, Justin
and Lee, Joonho
and Lee, Kenny
and Lensky, Yuri D.
and Leon, Shannon
and Lester, Brian J.
and Li, Wing Yan
and Li, Yin
and Lill, Alexander T.
and Liu, Wayne
and Livingston, William P.
and Locharla, Aditya
and Lucero, Erik
and Lundahl, Daniel
and Lunt, Aaron
and Madhuk, Sid
and Malone, Fionn D.
and Maloney, Ashley
and Mandr{\`a}, Salvatore
and Manyika, James
and Martin, Leigh S.
and Martin, Orion
and Martin, Steven
and Maxfield, Cameron
and McClean, Jarrod R.
and McEwen, Matt
and Meeks, Seneca
and Megrant, Anthony
and Mi, Xiao
and Miao, Kevin C.
and Mieszala, Amanda
and Molavi, Reza
and Molina, Sebastian
and Montazeri, Shirin
and Morvan, Alexis
and Movassagh, Ramis
and Mruczkiewicz, Wojciech
and Naaman, Ofer
and Neeley, Matthew
and Neill, Charles
and Nersisyan, Ani
and Neven, Hartmut
and Newman, Michael
and Ng, Jiun How
and Nguyen, Anthony
and Nguyen, Murray
and Ni, Chia-Hung
and Niu, Murphy Yuezhen
and O'Brien, Thomas E.
and Oliver, William D.
and Opremcak, Alex
and Ottosson, Kristoffer
and Petukhov, Andre
and Pizzuto, Alex
and Platt, John
and Potter, Rebecca
and Pritchard, Orion
and Pryadko, Leonid P.
and Quintana, Chris
and Ramachandran, Ganesh
and Reagor, Matthew J.
and Redding, John
and Rhodes, David M.
and Roberts, Gabrielle
and Rosenberg, Eliott
and Rosenfeld, Emma
and Roushan, Pedram
and Rubin, Nicholas C.
and Saei, Negar
and Sank, Daniel
and Sankaragomathi, Kannan
and Satzinger, Kevin J.
and Schurkus, Henry F.
and Schuster, Christopher
and Senior, Andrew W.
and Shearn, Michael J.
and Shorter, Aaron
and Shutty, Noah
and Shvarts, Vladimir
and Singh, Shraddha
and Sivak, Volodymyr
and Skruzny, Jindra
and Small, Spencer
and Smelyanskiy, Vadim
and Smith, W. Clarke
and Somma, Rolando D.
and Springer, Sofia
and Sterling, George
and Strain, Doug
and Suchard, Jordan
and Szasz, Aaron
and Sztein, Alex
and Thor, Douglas
and Torres, Alfredo
and Torunbalci, M. Mert
and Vaishnav, Abeer
and Vargas, Justin
and Vdovichev, Sergey
and Vidal, Guifre
and Villalonga, Benjamin
and Heidweiller, Catherine Vollgraff
and Waltman, Steven
and Wang, Shannon X.
and Ware, Brayden
and Weber, Kate
and Weidel, Travis
and White, Theodore
and Wong, Kristi
and Woo, Bryan W. K.
and Xing, Cheng
and Yao, Z. Jamie
and Yeh, Ping
and Ying, Bicheng
and Yoo, Juhwan
and Yosri, Noureldin
and Young, Grayson
and Zalcman, Adam
and Zhang, Yaxing
and Zhu, Ningfeng
and Zobrist, Nicholas
and AI, Google Quantum
and {Collaborators}},
title={Quantum error correction below the surface code threshold},
journal={Nature},
year={2025},
month={Feb},
day={01},
volume={638},
number={8052},
pages={920-926},
abstract={Quantum error correction1--4 provides a path to reach practical quantum computing by combining multiple physical qubits into a logical qubit, in which the logical error rate is suppressed exponentially as more qubits are added. However, this exponential suppression only occurs if the physical error rate is below a critical threshold. Here we present two below-threshold surface code memories on our newest generation of superconducting processors, Willow: a distance-7 code and a distance-5 code integrated with a real-time decoder. The logical error rate of our larger quantum memory is suppressed by a factor of $\Lambda${\thinspace}={\thinspace}2.14{\thinspace}{\textpm}{\thinspace}0.02 when increasing the code distance by 2, culminating in a 101-qubit distance-7 code with 0.143{\%}{\thinspace}{\textpm}{\thinspace}0.003 per cent error per cycle of error correction. This logical memory is also beyond breakeven, exceeding the lifetime of its best physical qubit by a factor of 2.4{\thinspace}{\textpm}{\thinspace}0.3. Our system maintains below-threshold performance when decoding in real time, achieving an average decoder latency of 63{\thinspace}microseconds at distance 5 up to a million cycles, with a cycle time of 1.1{\thinspace}microseconds. We also run repetition codes up to distance 29 and find that logical performance is limited by rare correlated error events, occurring approximately once every hour or 3{\thinspace}{\texttimes}{\thinspace}109 cycles. Our results indicate device performance that, if scaled, could realize the operational requirements of large-scale fault-tolerant quantum algorithms.},
issn={1476-4687},
doi={10.1038/s41586-024-08449-y},
url={https://doi.org/10.1038/s41586-024-08449-y}
}

@article{PhysRevA.91.022118,
  title = {Tomography via correlation of noisy measurement records},
  author = {Ryan, Colm A. and Johnson, Blake R. and Gambetta, Jay M. and Chow, Jerry M. and da Silva, Marcus P. and Dial, Oliver E. and Ohki, Thomas A.},
  journal = {Phys. Rev. A},
  volume = {91},
  issue = {2},
  pages = {022118},
  numpages = {7},
  year = {2015},
  month = {Feb},
  publisher = {American Physical Society},
  doi = {10.1103/PhysRevA.91.022118},
  url = {https://link.aps.org/doi/10.1103/PhysRevA.91.022118}
}

@article{PhysRevA.52.R2493,
  title = {Scheme for reducing decoherence in quantum computer memory},
  author = {Shor, Peter W.},
  journal = {Phys. Rev. A},
  volume = {52},
  issue = {4},
  pages = {R2493(R)--R2496(R)},
  numpages = {0},
  year = {1995},
  month = {Oct},
  publisher = {American Physical Society},
  doi = {10.1103/PhysRevA.52.R2493},
  url = {https://link.aps.org/doi/10.1103/PhysRevA.52.R2493}
}

@article{Kitaev_1997,
doi = {10.1070/RM1997v052n06ABEH002155},
url = {https://doi.org/10.1070/RM1997v052n06ABEH002155},
year = {1997},
month = {dec},
publisher = {},
volume = {52},
number = {6},
pages = {1191},
author = {A Yu Kitaev},
title = {Quantum computations: algorithms and error correction},
journal = {Russian Mathematical Surveys}
}

@Article{Campbell2017,
author={Campbell, Earl T.
and Terhal, Barbara M.
and Vuillot, Christophe},
title={Roads towards fault-tolerant universal quantum computation},
journal={Nature},
year={2017},
month={Sep},
day={01},
volume={549},
number={7671},
pages={172-179},
issn={1476-4687},
doi={10.1038/nature23460},
url={https://doi.org/10.1038/nature23460}
}

@article{montanaro2016quantum,
  title={Quantum algorithms: an overview},
  author={Montanaro, Ashley},
  journal={npj Quantum Information},
  volume={2},
  number={1},
  pages={15023},
  year={2016},
  publisher={Nature Publishing Group}
}

@article{cramer2016repeated,
  title={Repeated quantum error correction on a continuously encoded qubit by real-time feedback},
  author={Cramer, Julia and Kalb, Norbert and Rol, M Adriaan and Hensen, Bas and Blok, Machiel S and Markham, Matthew and Twitchen, Daniel J and Hanson, Ronald and Taminiau, Tim H},
  journal={Nature communications},
  volume={7},
  number={1},
  pages={11526},
  year={2016},
  publisher={Nature Publishing Group UK London}
}

\end{document}